# The direct force correction based framework for general co-rotational analysis

Ziyun Kan[1,2]*, Kaijun Dong[1], Biaosong Chen[1], Haijun Peng[1]*, Xueguan Song[2]

1 State Key Laboratory of Structural Analysis for Industrial Equipment, Department of Engineering Mechanics, Dalian University of Technology, Dalian 116024, People 's Republic of China.

2 School of Mechanical Engineering, Dalian University of Technology, Dalian 116024, People's Republic of China.

## Abstract

The use of nonlinear projection matrix in co-rotational (CR) analysis was pioneered by Rankin and Nour-Omid in 1990s (*Computers & Structures*, 30 (1988) 257-267; *Comput. Methods Appl. Mech. Engrg*., 93 (1991) 353-384), and has almost became a standard manner for CR formulations deduction over the past thirty years. This matrix however relies heavily on a hysterical and sophisticated derivation of the variation of the local displacements to the global ones, leading to complicated expressions for the internal force vector and the tangent stiffness matrix, which may devalue the simplicity and convenience for the original intention of using CR approach. This paper begins by making a discussion on existing element independent CR formulation and the objective is to develop a new and simple framework for general CR analysis that avoids using conventional nonlinear projection matrix. The methodology consists of two steps in the element calculation. The first one is to obtain a preliminary result of the internal force. This is done by following the conventional element-independent CR formulation but drops the terms involving projection matrix and therefore yields simple formulations of the internal force and the tangent stiffness matrix. The second one is a correction step to obtain a new internal force vector that satisfies the element self-equilibrium condition. This step inherits the spirt of using projection matrix but is conducted directly in the global frame, thus avoiding complicated entanglement of local-global rotation and is independent of the choice of the local CR frame used in the CR analysis. This further leads to a simple and unified formulation for different kinds of elements that can be cooperated in CR framework. Closed formulation of the correction force as well as the related consistent tangent stiffness matrix are derived for different correction approaches. It is also shown that the existing linear projector matrix used for infinitesimal rotation analysis is a special case of the current correction approaches. Multiple numerical examples involving various kinds of elements and different choices of element local CR frame are presented to demonstrate the performance of the proposed framework. The outcomes show that for all the examples the accuracy of the results are comparable with those obtained in conjunction with conventional nonlinear projection matrix.

---

* Corresponding author

Email: kanziyun@ dlut.edu.cn (Ziyun Kan); hjpeng@dlut.edu.cn (Haijun Peng)

## 1. Introduction

Geometrically nonlinear analysis of structures is frequently encountered in a variety of engineering applications. The total Lagrangian and updated Lagrangian are the earliest developed finite element formulations to deal with this kind of analysis. Alternative to these two Lagrangian formulations, the co-rotational (CR) approach is viewed as a simple way to conduct nonlinear finite elements analysis for large displacements but small strain problems, and generated an increased amount of interest in the last decades. It is largely recognized that CR approach was first introduced by Wempner [1], latter contributed by Belytschko and co-workers [2, 3], and has much in common with the “natural approach” proposed by Argyris et al. [4]. The main concept of CR approach is to decompose the motion of the element into rigid body motion and the deformational parts, through the use of a reference system which continuously rotates with the element. The deformational response is captured at the level of the element local frame, whereas the geometric nonlinearity induced by the large rigid-body motion, is incorporated in the transformation matrices relating local and global quantities.

In the development history of CR approach, important contributions were made by Rankin and co-workers [5-7], who developed an “element independent CR formulation”. This formulation allows upgrading existing high-performance linear finite elements for CR analysis in a systematic process. The element independent CR formulation relies heavily on the use of projection matrix [6, 7], which is a product of the variation of the local displacements to the variation of the global ones, and is closely related to the choice of the local CR frame. This variation emerges in the stage of deriving the internal force vector. A twice variation would therefore be needed to obtain a consistent tangent stiffness matrix. The use of projection matrix is believed to be able to extract the “pure” deformation part to the global displacements and to ensure the self-equilibrium property of the element. It is shown in [6, 7] that the performance of quadrilateral warping shell element with coarse mesh was significantly improved with projection matrix. After then, the using of projection matrix was soon extended to a large variety of CR analyses involving various kinds of elements, such as plane [8-10], solid/solid-shell [11-14], beam [15-21], triangular shell [22-33] and quadrilateral shell [34-36] elements. In some works [37], the projection matrix is not explicitly stated, but is essentially derived in the formulation since the full variation of the local displacements to the global ones is conducted in the process of internal force calculation. Apart from Rankin and Nour-Omid’s pioneering works [6, 7], it appears that the necessity of using projector matrix in these existing works was less investigated. In most of these existing works, the authors simply deduce element formulations follow a standard manner and emphasize the validity of their formulation through numerical examples. Few attention was put on the issue that to what extent the projector matrix embodied in their formulations can enhance the performance of the specific element they investigated. A particular reference should be mentioned was given by Felippa and Haugen [38], who provide an excellent review on the element independent CR framework.

The use of projection matrix can surely improve the element performance for CR

analysis. This is done, however, through a hysterical and sophisticated derivation of the variations of the local displacements to the variations of the global ones, which leads to a rather complicated expression for the internal force vector, as well the tangent stiffness matrix and may devalue the simplicity and convenience for the original intention of using CR approach. This is in particular for the case if the local CR frame is defined by some indirect rule such as the polar-decomposition based approach [25, 39], and it is hard, if not impossible, to obtain an explicit expression for the derivation of the projection matrix. The problem became even challenging for the derivation of tangent stiffness matrix, since a further variation of the projection matrix is needed. In most of the exiting works stating that a consistent tangent stiffness matrix is calculated, some terms regarding to the variation of the projection matrix was actually dropped given its complexity. This is even for the case of using the simplest choice of the element local co-rational frame: the side alignment approach.

The objective of this work is to develop a new and simple framework for general CR analysis. The methodology proposed here inherit the spirt of the element independent CR formulation derived by Rankin and co-workers but avoid the entanglement of the complex derivation of the nonlinear projector matrix in the stage of force calculation. The methodology consists of two steps in the element calculation. The first one is to obtain a preliminary result of the internal force by dropping the terms involving projection matrix. We remark this step essentially follows with the earliest version of CR formulation. A simple formulation of the internal force and the consistent tangent stiffness matrix would therefore be easily obtained. The second one is a direct force correction step based on the preliminary internal force vector to obtain a new internal force vector that satisfies the element self-equilibrium. The correction step is conducted directly in the global frame, thus avoiding complicated entanglement of local-global transformation and is independent of the choice of the element local CR frame. This further leads to a simple and unified formulation for different kinds of elements that can be cooperated in CR framework. Closed formulations of the correction force term as well as the related tangent stiffness matrix are rigorously derived. The connection of the proposed force correction approaches with existing linear projection matrix are also revealed. A number of numerical examples involving various kinds of elements and different choices of local CR frame are presented. Emphasis is placed on comparing the results obtained by the proposed approaches with those obtained by formulations with conventional projection matrix. The enhancement of using conventional nonlinear projection matrix for different examples are also revealed, which seems seldom be addressed.

## 2. Discussions on existing element-independent CR formulation

In this section, we review and discuss the existing commonly used element-independent CR formulation, which is closely related to our following work. We provide some new insights into this approach. Some simplifications in deriving the formulation are also given. To best highlight the core concept of CR framework and avoid involving too much about complex 3D finite rotation involving rotational degrees of freedom, we first limit our discussions to continuum elements. Discussion on structural elements with rotational degrees of freedom will be given subsequently. Throughout this work, we limit our analysis to linear elastic static case, although the formulations may potentially be extended to broader cases.

## 2.1 Continuum element

As shown in Fig. 1, we consider a general continuum element with $N$ nodes, with each node consisting of three translational degrees of freedom. $OXYZ$ denotes the global coordinate frame. The element stiffness matrix associated with global frame is denoted as $\boldsymbol{K} \in \Re^{3N\times 3N}$, which is a symmetric positive semidefinite matrix. If the element is subjected to only a small nodal displacement $\boldsymbol{u} \in \Re^{3N}$ such that the element rotation is small, the strain energy can be given as $\Phi(\boldsymbol{u}) = \boldsymbol{u}^{\mathrm{T}}\boldsymbol{K}\boldsymbol{u}/2 \in \Re$. Trivially, by taking the variation of the strain energy to the displacement vector the nodal internal force vector can be obtained as $\boldsymbol{f} = \left[\mathrm{d}\Phi/\mathrm{d}\boldsymbol{u}\right]^{\mathrm{T}} = \boldsymbol{K}\boldsymbol{u} \in \Re^{3N}$. The physical meaning of the element internal force vector is the nodal translational forces measured in the global frame. The element stiffness matrix has the property that for arbitrary rigid body translation of the element $\boldsymbol{u}_{\mathrm{tran}} = [\boldsymbol{d}^{\mathrm{T}}, \boldsymbol{d}^{\mathrm{T}}, \cdots, \boldsymbol{d}^{\mathrm{T}}]^{\mathrm{T}} \in \Re^{3N}$, where $\boldsymbol{d} \in \Re^{3}$ is the translation of every material point at the element, no internal force will be produced (i.e., $\boldsymbol{K}\boldsymbol{u}_{\mathrm{tran}} = \boldsymbol{0}$).

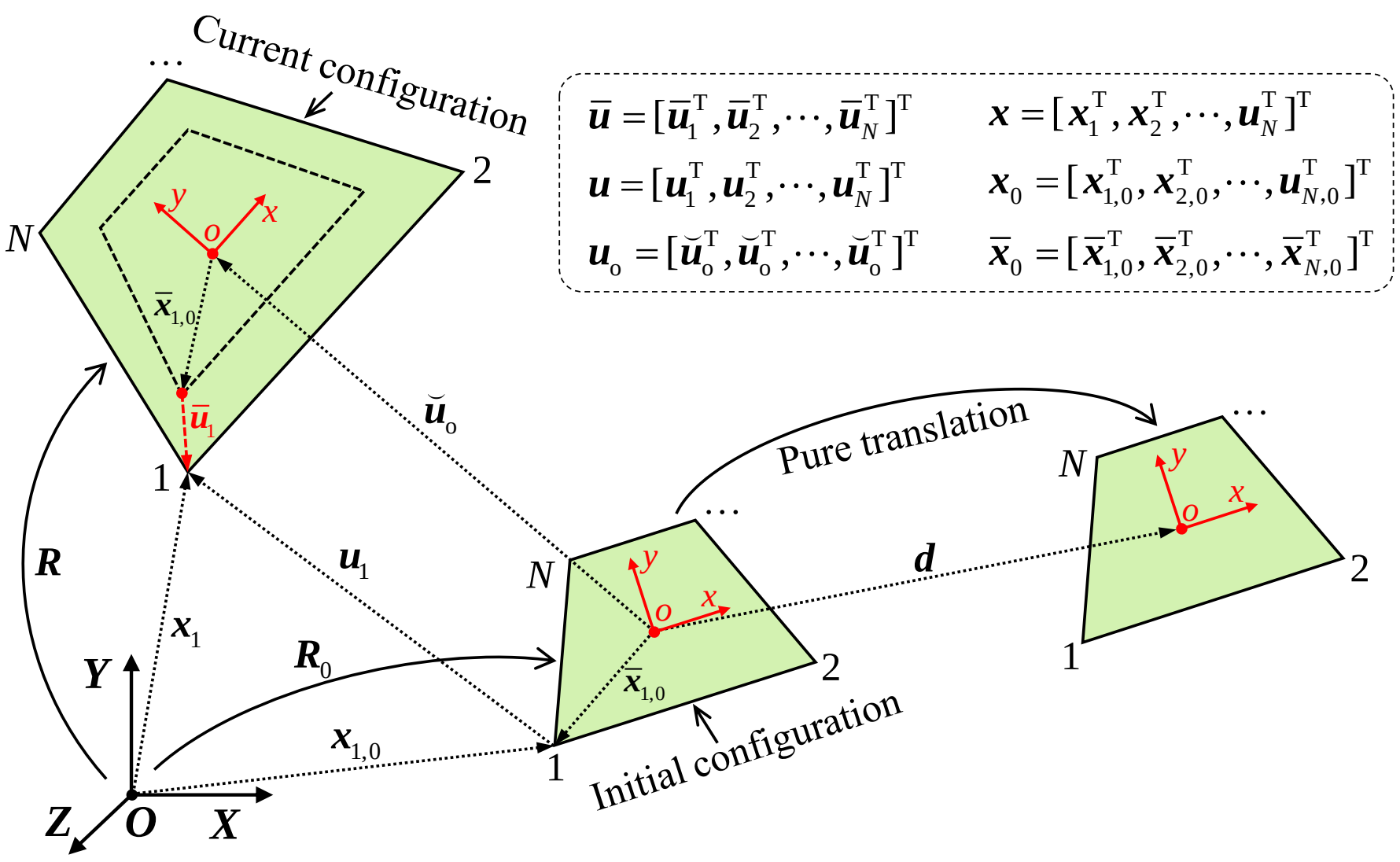

Fig. 1 Co-rotational formulation of continuum element: kinematics and coordinate systems.

We now consider the case that element is subjected to a large nodal displacement. The main idea of the CR approach is to extract the local deformation part from the large nodal displacement, such that the strain energy function can still be approximated at the small deformation region. This is done through defining a local CR coordinate system and measuring all deformation terms in this local CR frame. With the help of element local CR frame, the element kinematics can be split into two steps. The first step is a rigid translation and rotation of the local CR frame, and the second one is a local deformation with respect to this local CR frame. The orientation of the element local CR frame at the initial configuration (specified by the orthogonal matrix $\boldsymbol{R}_0 \in \Re^{3\times 3}$) can be defined arbitrarily (it is typically chosen such that one of the axis are parallel to one side of the element), as long as the element local stiffness matrix is consistent with the defined local CR frame. We denote the element local stiffness matrix as $\bar{\boldsymbol{K}}$ (throughout this paper, an overbar means that the term is measured with respect with the local CR frame). Following many existing works [7], the local deformational displacement of the element, $\bar{\boldsymbol{u}} \in \Re^{3N}$, is given as

$$\bar{\boldsymbol{u}}\left(\boldsymbol{u}, \boldsymbol{R}(\boldsymbol{u})\right) = \mathrm{diag}\left(\boldsymbol{R}^{\mathrm{T}}\right)\left(\boldsymbol{u} + \boldsymbol{x}_0 - \boldsymbol{u}_{\mathrm{o}} - \boldsymbol{x}_{\mathrm{o},0}\right) - \bar{\boldsymbol{x}}_0 \tag{1}$$

where $\boldsymbol{u} \in \Re^{3N}$ and $\boldsymbol{x}_0 \in \Re^{3N}$ denote the global displacement and initial position of the nodes, respectively; $\bar{\boldsymbol{x}}_0 \in \Re^{3N}$ denotes the initial nodal position vector measured in the element local CR frame, which is a constant throughout the analysis. $\mathrm{diag}\left(\boldsymbol{R}^{\mathrm{T}}\right) \in \Re^{3N\times 3N}$ denotes a block diagonal matrix composed of the transpose of the orthogonal matrix $\boldsymbol{R} \in \Re^{3\times 3}$. The orthogonal matrix representing the rigid rotation of the element local CR frame, which is evaluated using some direct or indirect rules based on the current configuration and serves as an important quantity to remove the element rigid rotation. In Eq. (1), $\boldsymbol{u}_{\mathrm{o}} \in \Re^{3N}$ and $\boldsymbol{x}_{\mathrm{o}} \in \Re^{3N}$ are the column vector arrayed by the global displacement and initial position of the origin of the local CR frame. The origin of the local CR frame is typically attached to one node or the central point of the element. We should, however, remark that the origin can be defined arbitrarily. Different choices of the origin only involve a rigid translation of the element and this would not influence the calculation of the local internal force. In many works, the core concept of the CR approach is stated as to define a local coordinate system that continuously “rotates" and “translates" with the element. The above fact implies that the “rotates" is essential while the “translates" is unimportant. This is perhaps the reason why the formulation is referred to as the co-“rotational”, instead of the co-“rotational and translational” formulation or some similar terminologies. Therefore, in the following discussions we only focus on the influence of the rotation on the analysis.

In the above sense, formulation (1) can be simplified as

$$\bar{\boldsymbol{u}}\left(\boldsymbol{u}, \boldsymbol{R}(\boldsymbol{u})\right) = \mathrm{diag}\left(\boldsymbol{R}^{\mathrm{T}}\right)\boldsymbol{x} - \bar{\boldsymbol{x}}_0 \tag{2}$$

where $\boldsymbol{x} = \boldsymbol{u} + \boldsymbol{x}_0 \in \Re^{3N}$ denote the current position vector of the element nodes.

**Remark**: In some works [8, 14], the element rigid rotation is determined by using a least square fit of the local deformational displacement (i.e, $\boldsymbol{R} = \arg\min\left(\bar{\boldsymbol{u}}^{\mathrm{T}}\bar{\boldsymbol{u}}\right)$ ). The origin of the element local CR frame should be considered in the determination of the element rotation, and the optimization problem leads to a solution that the origin must be placed at the element centroid position. Once $\boldsymbol{R}$ is calculated, expression (2) is still applicable for the internal force calculation.

Based on the core concept of CR, the strain energy function is obtained as

$$\Phi(\boldsymbol{u}) = \bar{\boldsymbol{u}}(\boldsymbol{u})^{\mathrm{T}} \bar{\boldsymbol{K}}\bar{\boldsymbol{u}}(\boldsymbol{u}) / 2 \tag{3}$$

The global nodal internal force vector is obtained by taking the derivative of the strain energy to the global displacement vector, as

$$\boldsymbol{f}(\boldsymbol{u}) = \left[\frac{\mathrm{d}\Phi}{\mathrm{d}\boldsymbol{u}}\right]^{\mathrm{T}} = \left[\frac{\mathrm{d}\Phi}{\mathrm{d}\bar{\boldsymbol{u}}}\frac{\mathrm{d}\bar{\boldsymbol{u}}}{\mathrm{d}\boldsymbol{u}}\right]^{\mathrm{T}} = \left[\frac{\mathrm{d}\bar{\boldsymbol{u}}}{\mathrm{d}\boldsymbol{u}}\right]^{\mathrm{T}} \bar{\boldsymbol{K}}\bar{\boldsymbol{u}} \tag{4}$$

where the term $\bar{\boldsymbol{K}}\bar{\boldsymbol{u}}$ denotes the force vector expressed in the local CR frame, and serves as a transformation matrix between the local force and the global one. The element consistent tangent stiffness matrix can be further obtained as the derivative of global nodal internal force vector to the global displacement vector, as $\boldsymbol{K} = \mathrm{d}\boldsymbol{f} / \mathrm{d}\boldsymbol{u}$.

**Remark**: The above deduction from a strain energy view point essentially coincides with many existing works [8, 38] that adopt the principle of virtual work to derive the CR formulations. The merits of the deduction in this work are twofold. First, one can see that the obtained consistent tangent stiffness matrix is a Hessian matrix of the strain energy function $\Phi$, therefore it must be symmetrical for a continuum element regardless how the local CR frame is chosen in the analysis. It appears that the statements in [11] that a non-symmetric tangent stiffness matrix would be obtained for three-dimensional continuum elements is erroneous. Second, this kind of deduction allows a geometrical insight into the using of conventional projection matrix in CR analysis, as well as an insight into how different choices of local CR frame influence the internal force results.

With Eq. (2), the derivation of the transformation matrix can be split into two parts

$$\frac{\mathrm{d}\bar{\boldsymbol{u}}}{\mathrm{d}\boldsymbol{u}} = \mathrm{diag}\left(\boldsymbol{R}^{\mathrm{T}}\right) + \left.\frac{\partial\left(\mathrm{diag}\left(\boldsymbol{R}^{\mathrm{T}}\right)\boldsymbol{x}\right)}{\partial\boldsymbol{u}}\right|_{\boldsymbol{x}} \tag{5}$$

where the first part represents the ordinary orthogonal transformation from the local CR frame to the global one; the second part is derived by keeping $\boldsymbol{x}$ constant (in the following, the notation $(\bullet)|_{\boldsymbol{x}}$ will be omitted for brevity), and it reflects the variation of the orthogonal matrix in response to the variation of the nodal displacements. Rankin and Nour-Omid [6, 7] made use of the instantaneous axial vector to obtain the variation of the orthogonal matrix. Following their methodology, the second term of the transformation matrix (5) can be derived as (see in the Appendix)

$$\frac{\partial\left(\mathrm{diag}\left(\boldsymbol{R}^{\mathrm{T}}\right)\boldsymbol{x}\right)}{\partial\boldsymbol{u}} = -\bar{\boldsymbol{S}}\bar{\boldsymbol{G}}\mathrm{diag}\left(\boldsymbol{R}^{\mathrm{T}}\right) \tag{6}$$

where $\bar{\boldsymbol{S}} \in \Re^{3N\times 3}$ is the spin-lever or moment-arm matrix [38], and for continuum element it is defined as

$$\bar{\boldsymbol{S}} = \left[\operatorname{spin}(\bar{\boldsymbol{x}}_1) \quad \operatorname{spin}(\bar{\boldsymbol{x}}_2) \quad \cdots \quad \operatorname{spin}(\bar{\boldsymbol{x}}_N)\right]^{\mathrm{T}} \tag{7}$$

The operation "spin" above is related to the cross-product and takes the following form for a $3\times 1$ vector $\boldsymbol{r} = [r_1, r_2, r_3]^{\mathrm{T}}$

$$\operatorname{spin}(\boldsymbol{r}) = \boldsymbol{r}\times = \begin{bmatrix} 0 & -r_3 & r_2 \\ r_3 & 0 & -r_1 \\ -r_2 & r_1 & 0 \end{bmatrix} = -\operatorname{spin}(\boldsymbol{r})^{\mathrm{T}} \tag{8}$$

Matrix $\bar{\boldsymbol{G}} \in \Re^{3\times 3N}$ in Eq. (6) is the spin-fitter matrix [38], which links the variation of the element instantaneous spin axial vector in response to the variation of the local nodal displacements:

$$\bar{\boldsymbol{G}} = \begin{bmatrix} \dfrac{\partial \bar{\boldsymbol{\omega}}}{\partial \bar{\boldsymbol{u}}_1} & \dfrac{\partial \bar{\boldsymbol{\omega}}}{\partial \bar{\boldsymbol{u}}_2} & \cdots & \dfrac{\partial \bar{\boldsymbol{\omega}}}{\partial \bar{\boldsymbol{u}}_N} \end{bmatrix} \tag{9}$$

where $\bar{\boldsymbol{\omega}}$ denotes the instantaneous axial vector of the orthogonal transformation expressed in the local CR frame. Detailed expression of matrix $\bar{\boldsymbol{G}}$ depends on how to approximate the orthogonal matrix $\boldsymbol{R}$ based on the current configuration, and it is typically a highly nonlinear function of the current quantities. A further variation of this term is needed to obtain the consistent tangent stiffness matrix, and is, however, neglected in many existing works given its complexity. As argued by Felippa and Haugen [38] , this may be not appropriate for some occasions such as that with highly warped shell elements in a coarse mesh. Combining Eqs. (4)~(6), the element global internal force vector can be expressed as

$$\begin{aligned} \boldsymbol{f}(\boldsymbol{u}) &= \operatorname{diag}(\boldsymbol{R})\left(\boldsymbol{I}_{3N} - \bar{\boldsymbol{S}}\bar{\boldsymbol{G}}\right)^{\mathrm{T}} \bar{\boldsymbol{K}}\bar{\boldsymbol{u}} \\ &= \operatorname{diag}(\boldsymbol{R})\bar{\boldsymbol{K}}\bar{\boldsymbol{u}} - \operatorname{diag}(\boldsymbol{R})\bar{\boldsymbol{G}}^{\mathrm{T}}\bar{\boldsymbol{S}}^{\mathrm{T}}\bar{\boldsymbol{K}}\bar{\boldsymbol{u}} \end{aligned} \tag{10}$$

where $\boldsymbol{I}_{3N}$ is the $3N\times 3N$ identity matrix; the term $\boldsymbol{I}_{3N} - \bar{\boldsymbol{S}}\bar{\boldsymbol{G}}$ is typically denoted as $\bar{\boldsymbol{P}}$, representing the "*projection matrix*". For easy of discussion, we split the expression into two terms. The first term, $\operatorname{diag}(\boldsymbol{R})\bar{\boldsymbol{K}}\bar{\boldsymbol{u}}$, represents the ordinary orthogonal transformation between the local force to the global one and we should remark that it is the *main part* of the internal force, while the second term, $\operatorname{diag}(\boldsymbol{R})\bar{\boldsymbol{G}}^{\mathrm{T}}\bar{\boldsymbol{S}}^{\mathrm{T}}\bar{\boldsymbol{K}}\bar{\boldsymbol{u}}$, is a small quantity representing a correction term to the main part. The effects of this correction is to turn the moment-unbalanced force into a balanced one. To illustrate this, we focus on the force at the local level. It should be first noted that the product $\bar{\boldsymbol{S}}^{\mathrm{T}}\bar{\boldsymbol{K}}\bar{\boldsymbol{u}}$ is simply the unbalanced moment produced by the local force (i.e, $\bar{\boldsymbol{S}}^{\mathrm{T}}\bar{\boldsymbol{K}}\bar{\boldsymbol{u}} \equiv \sum_{i=1}^{N} \bar{\boldsymbol{x}}_i \times \bar{\boldsymbol{f}}_i$ ,where $\bar{\boldsymbol{f}}_i$ is the local translational force of a particular node *i*). Rankin and Nour-Omid [6, 7] show that for certain choices of element local CR frame, the bi-orthogonality relationship, $\bar{\boldsymbol{S}}^{\mathrm{T}}\bar{\boldsymbol{G}}^{\mathrm{T}} = \bar{\boldsymbol{G}}\bar{\boldsymbol{S}} = \boldsymbol{I}_3$, is satisfied. As a result, we have $\bar{\boldsymbol{S}}^{\mathrm{T}}\left(\boldsymbol{I}_{3N} - \bar{\boldsymbol{S}}\bar{\boldsymbol{G}}\right)^{\mathrm{T}}\bar{\boldsymbol{K}}\bar{\boldsymbol{u}} \equiv \boldsymbol{0}$, which means that the original local force ($\bar{\boldsymbol{K}}\bar{\boldsymbol{u}}$) is moment-balanced after adding the correction term ($-\bar{\boldsymbol{G}}^{\mathrm{T}}\bar{\boldsymbol{S}}^{\mathrm{T}}\bar{\boldsymbol{K}}\bar{\boldsymbol{u}}$). Matrix $\bar{\boldsymbol{G}}$ serves as an assignment matrix to equivalent the (local) unbalanced moment ($\bar{\boldsymbol{S}}^{\mathrm{T}}\bar{\boldsymbol{K}}\bar{\boldsymbol{u}}$) to the (local) translational force.

In CR analyses, the unbalanced moment produced by the internal force stems from the fact that the orthogonal matrix is typically evaluated using some direct or indirect rules based on the current configuration, and cannot remove the effect of element rigid-body rotation cleanly. In other words, for any given current/deformed configuration if there is rule to evaluate the orthogonal matrix (rigid-body rotation) such that no unbalanced moment is produced, there is completely no use to add the second term of Eq. (10) in the analysis (as it always maintain vanishing state in the global iteration, i.e., $\bar{\boldsymbol{S}}^{\mathrm{T}}\bar{\boldsymbol{K}}\bar{\boldsymbol{u}} \equiv \boldsymbol{0}$). For some special elements, such a rule exists and is simple:

(1) Two node straight bar element with the element rigid-body rotation being determined by the axis alignment approach.

(2) Three node constant strain triangle and four node constant strain tetrahedron elements with the element rigid-body rotation being determined by using the polar decomposition theorem of the deformation gradient.

These elements share some same features: the displacement field is a linear function of the nodal displacements; the rotation of each material point of the element is exactly the same, and therefore a single orthogonal matrix is sufficient to characterize the rigid-body rotation of the whole element; the correct rigid-body rotation corresponds to a condition that the curl of the local displacement field is zero.

We now consider an important theoretical question that for general elements under what choice of orthogonal matrix (rigid-body rotation) the second (force correction) term of Eq. (10), induced the variation of the orthogonal matrix, vanishes exactly.

**Lemma:** For any given current configuration $\boldsymbol{u}^*$, the force correction term vanishes exactly for the orthogonal matrix $\boldsymbol{R}$ is chosen such that the strain energy $\Phi\left(\boldsymbol{R}, \boldsymbol{u}^*\right)$ is located at the stationary point.

**Proof:** To prove this, it would be convenient to formally parameterize the orthogonal matrix using three parameter (such as Euler angel), i.e, $\boldsymbol{R}(\boldsymbol{u}) \to \boldsymbol{R}\left(\boldsymbol{\Theta}(\boldsymbol{u})\right)$. The local displacements can be reformulated as a function of the current displacement and the parameterized orthogonal matrix as $\bar{\boldsymbol{u}}\left(\boldsymbol{u}, \boldsymbol{\Theta}\right)$, and the second term in the (5) yields

$$\frac{\partial\left(\mathrm{diag}\left(\boldsymbol{R}^{\mathrm{T}}\right)\boldsymbol{x}\right)}{\partial\boldsymbol{u}}=\frac{\partial\left(\mathrm{diag}\left(\boldsymbol{R}(\boldsymbol{\Theta})^{\mathrm{T}}\right)\boldsymbol{x}\right)}{\partial\boldsymbol{\Theta}}\frac{\partial\boldsymbol{\Theta}}{\partial\boldsymbol{u}} \tag{11}$$

For any given deformed configuration, the element strain energy is only a function of the parameterized orthogonal matrix, i.e., $\Phi=\Phi(\boldsymbol{\Theta})=\bar{\boldsymbol{u}}(\boldsymbol{\Theta})^{\mathrm{T}}\bar{\boldsymbol{K}}\bar{\boldsymbol{u}}(\boldsymbol{\Theta})/2$ . The stationary point corresponds to a condition that

$$\frac{\mathrm{d}\Phi}{\mathrm{d}\boldsymbol{\Theta}}=\left[\frac{\partial\bar{\boldsymbol{u}}}{\partial\boldsymbol{\Theta}}\right]^{\mathrm{T}}\bar{\boldsymbol{K}}\bar{\boldsymbol{u}}=\left[\frac{\partial\left(\mathrm{diag}\left(\boldsymbol{R}(\boldsymbol{\Theta})^{\mathrm{T}}\right)\boldsymbol{x}\right)}{\partial\boldsymbol{\Theta}}\right]^{\mathrm{T}}\bar{\boldsymbol{K}}\bar{\boldsymbol{u}}\equiv\boldsymbol{0} \tag{12}$$

In combination with Eqs. (11) and (12), we can show that in this case the second term of the internal force vanishes exactly, i.e.,

$$\left[\frac{\partial\left(\mathrm{diag}\left(\boldsymbol{R}^{\mathrm{T}}\right)\boldsymbol{x}\right)}{\partial\boldsymbol{u}}\right]^{\mathrm{T}}\bar{\boldsymbol{K}}\bar{\boldsymbol{u}}=\left[\frac{\partial\boldsymbol{\Theta}}{\partial\boldsymbol{u}}\right]^{\mathrm{T}}\left[\frac{\partial\left(\mathrm{diag}\left(\boldsymbol{R}(\boldsymbol{\Theta})^{\mathrm{T}}\right)\boldsymbol{x}\right)}{\partial\boldsymbol{\Theta}}\right]^{\mathrm{T}}\bar{\boldsymbol{K}}\bar{\boldsymbol{u}}\equiv\boldsymbol{0} \tag{13}$$

To gain an insight into conventional CR analysis, a geometrical interpretation of the above deduction is presented by using Fig. 2. In the figure, the global displacement ($\boldsymbol{u}$) and the parameterized orthogonal matrix ($\boldsymbol{\Theta}$) are considered as independent variables, and the strain energy is plotted as a function of both the global displacement and the parameterized orthogonal matrix, i.e., $\Phi=\Phi(\boldsymbol{u},\boldsymbol{\Theta})$. The choice of the local CR frame artificially defines a constrained manifold between the global displacement and the orthogonal matrix, i.e., $\boldsymbol{\Theta}=\boldsymbol{\Theta}(\boldsymbol{u})$). The global iteration must be processed along the path defined by the intersection line between the strain energy function and the constrained manifold ($\Phi=\Phi(\boldsymbol{u},\boldsymbol{\Theta}(\boldsymbol{u}))$). The element internal force corresponds to the total derivative of the path with respect to the global displacement. For a particular global displacement $\boldsymbol{u}^{*}$, the force correction (second) term of Eq. (10) only when $\boldsymbol{\Theta}$ is chosen such that gradient of the strain energy function is perpendicular to the variation of the parameterized orthogonal matrix, which corresponds to a stationary point of the strain energy.

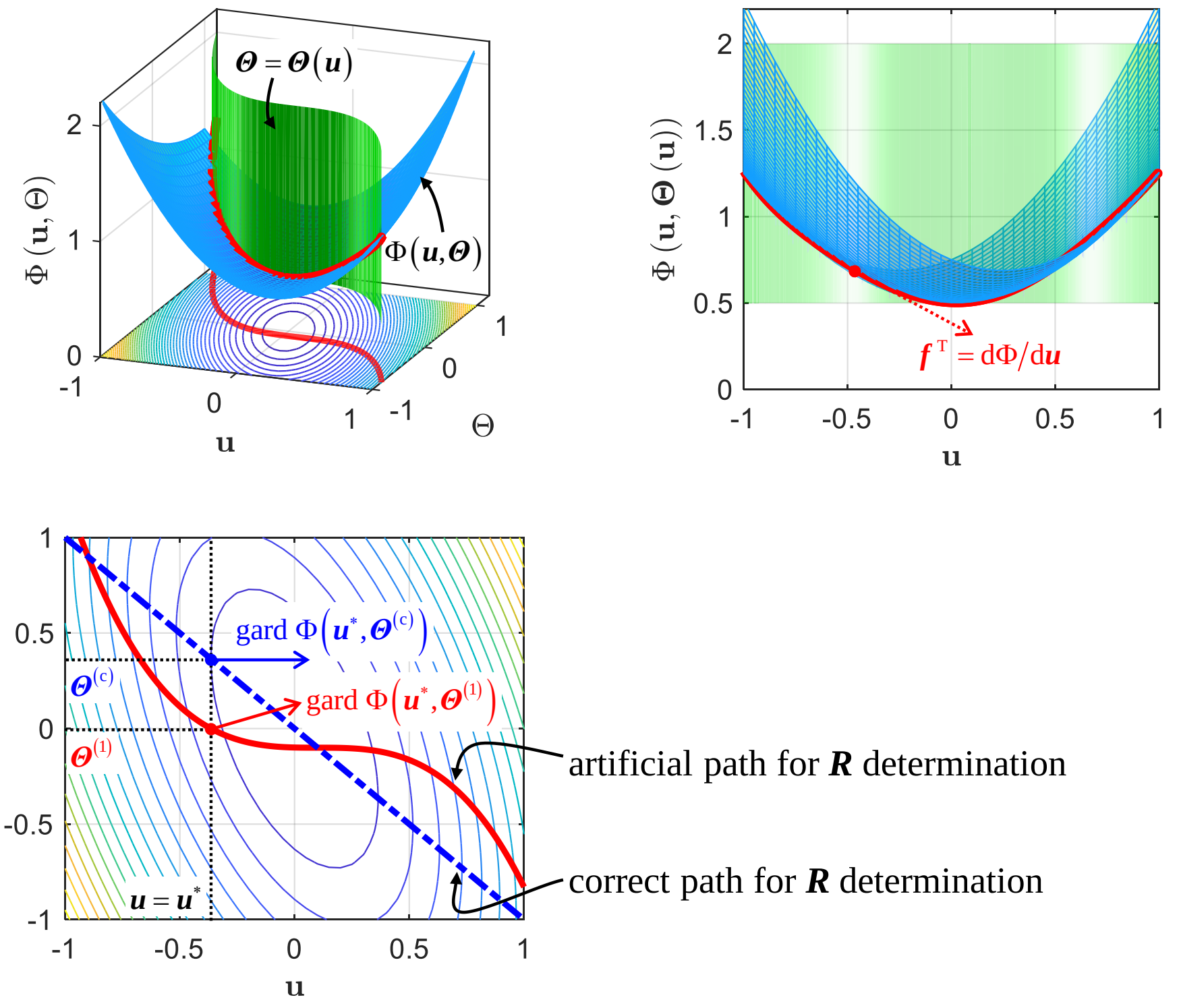

Fig. 2 Geometrical interpretation of the strain energy map and the influence of different choices of element local CR frame on the internal force calculation.

The above lemma opens an interesting question that is it possible to using the stationary condition in the CR analysis to enforce a correct rigid-body rotation for each element. This might be done through two approaches: the first one is adopting a mixed formulation CR analysis where both the global displacement and the parameterized orthogonal matrix are considered as primary variables in the global iteration and introducing the stationary condition as additional constraints imposed on the system; the second one follows with conventional displacement-based CR analysis, where the stationary condition is adopted to determine the element rotation for each current/updated/deformed configuration in the global iteration. (If the stationary point corresponds to the minimum point, the rotation matrix determination can be expressed as $\boldsymbol{R}(\boldsymbol{\Theta}) = \arg\min\left(\bar{\boldsymbol{u}}^{\mathrm{T}}\bar{\boldsymbol{K}}\bar{\boldsymbol{u}}/2\right)$, which interestingly shows a similarity with the existing least square fit approach by replacing the weight coefficient $\bar{\boldsymbol{K}}$ with $\boldsymbol{I}_{3N}$, and the solution may be obtained through an inner loop iteration). Unfortunately, according to the authors' tests, these approaches are only applicable for a limited number of elements within special deformation modes, and for general cases numerical unstable would occurs. This is due to the parasitic zero-energy mode in the element local stiffness matrix, and there are typically multiple stationary points of the strain energy function for an element under a deformed configuration. This fact can be revealed by using a simple two node plane straight bar element shown in Fig. 3. The bar is subjected to only a horizontal force, and it is clear that $\boldsymbol{\Theta} \equiv 0$ represents the correct rigid-body rotation, which always corresponds to stationary points of the strain energy (see in Fig. 4). However, if we follow the methodology of CR analysis, and intentionally consider an inappropriate choice of the local CR frame, we can find that for a particular current configuration if the bar is compressed ($u < 0$) there is only one stationary point, while if the bar is stretched ($u > 0$), at least three stationary points emerges. One represents the correct rigid-body rotation, and the other two correspond to the zero-energy modes (under which the local displacements are perpendicular to the axis of the CR configuration).

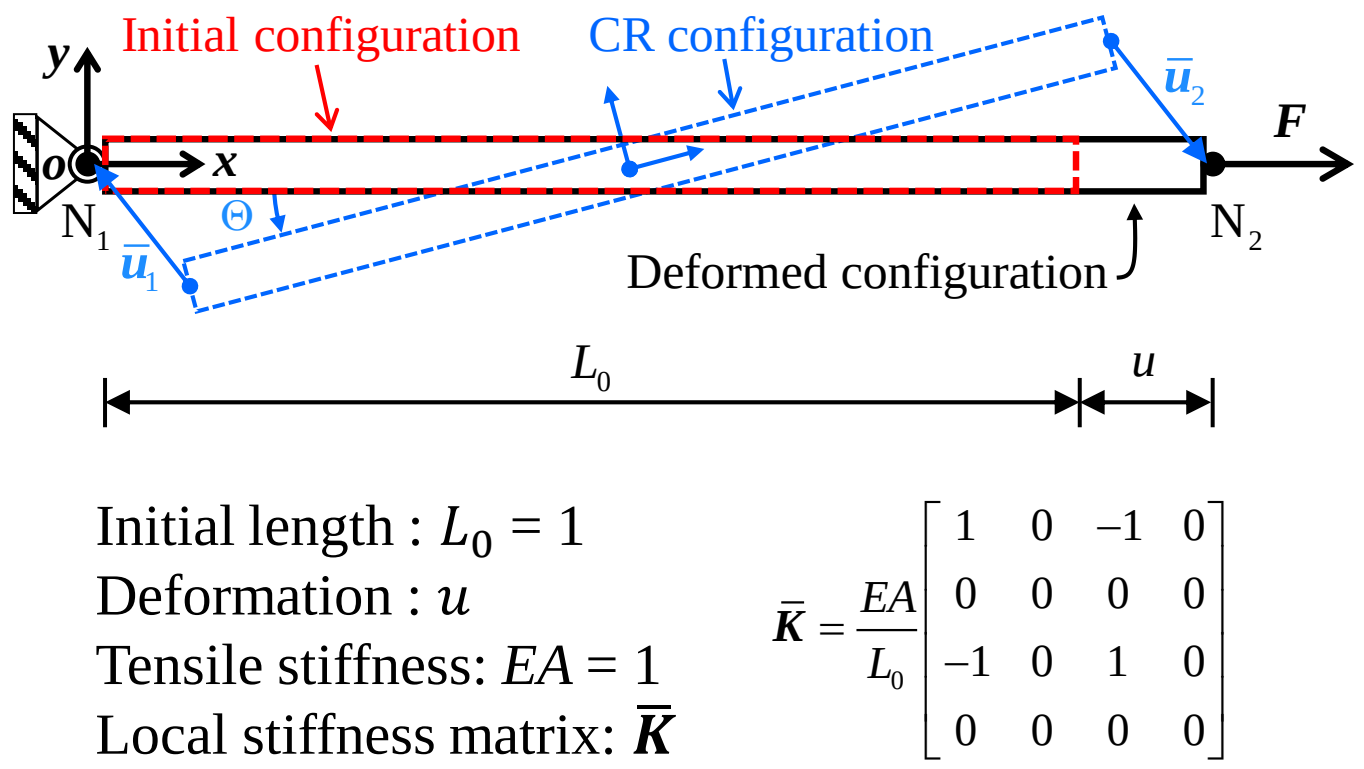

Fig. 3 A bar element that subjected to a horizontal force, with the CR approach being conducted for the analysis.

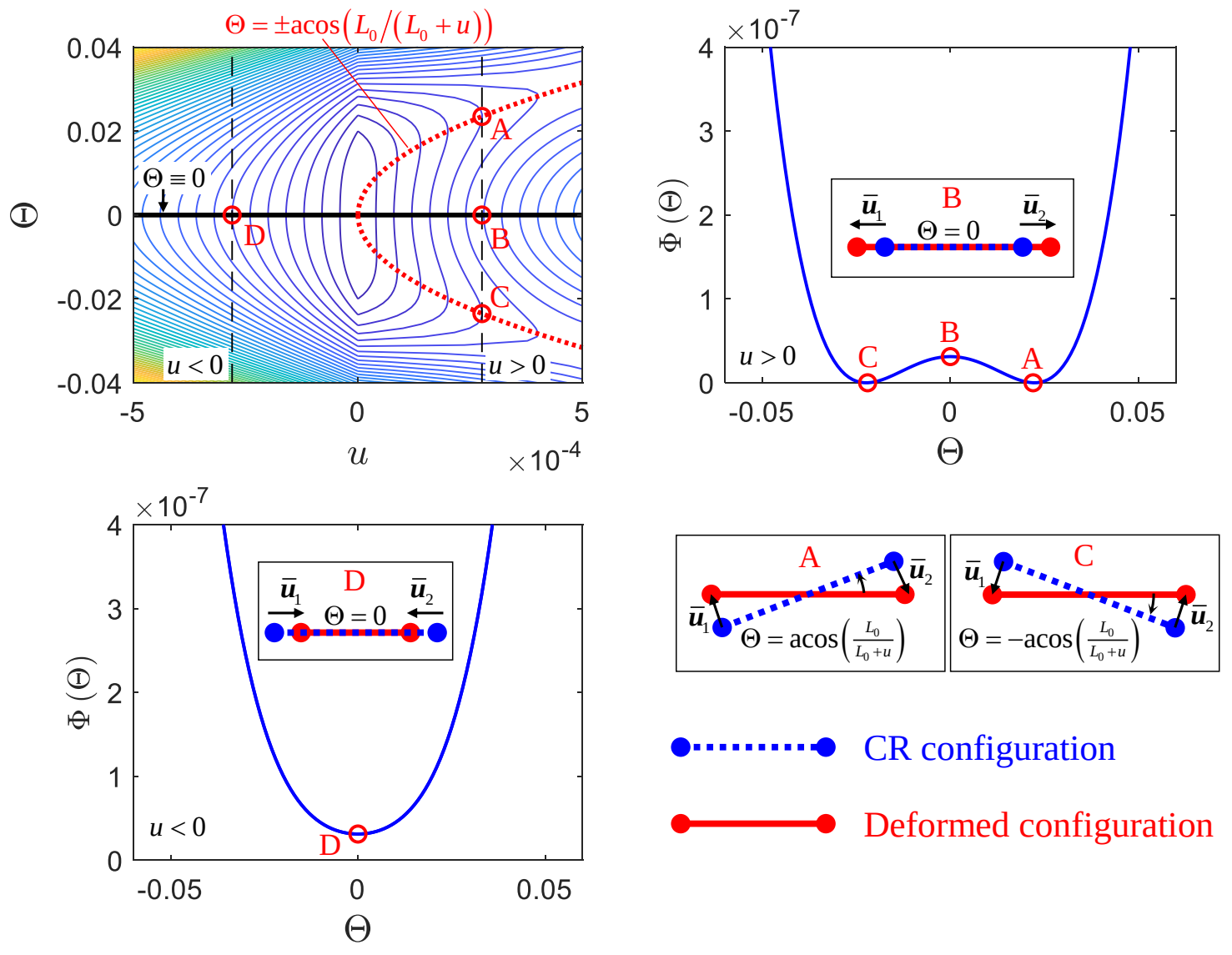

Fig. 4 Geometrical interpretation of the strain energy map for the bar element.

## 2.2 Structural element

The above discussions focus on continuum elements. For structural elements in CR analysis, the local deformation of the rotational freedoms should be considered. There are currently two different treatments regarding to how to deal with the rotational freedom. The first kind of treatment adopts rotational variables to parameterize the orthogonal rotation matrix associated with the nodal rotational freedom [16, 17, 22, 26, 35]. The merit of this kind of treatment is that the element strain energy function can still be expressed in terms of the parameterized variables as a quadratic form. As in the case of previous continuum elements, the tangent stiffness matrix of the internal force is still derived as the Hessian matrix of the strain energy function and is always symmetric. Besides, the rotational variables become additive and the necessity of a special updating procedure is avoided. The disadvantage of this approach is that the nodal internal force corresponding to the rotational freedoms cannot be interpreted as conventional accepted moment which leads to difficulties for applying external moment in the structural analysis. The second kind of treatment directly use an orthogonal rotation matrix to reflect the nodal rotation and use the instantaneous incremental spin variables to carry out the iteration. Due to the non-additivity of the instantaneous spin variables, the nodal rotation must be updated following the exponential mapping. One feature of this kind of treatment is that the element strain energy cannot be expressed in terms of these instantaneous incremental spin variables, and in general an asymmetric tangent stiffness matrix would be obtained in the global iterations excepted at the final convergent configuration. The advantage of this kind of treatment is that the internal force can be interpreted as the commonly accepted force and moments and would yields a relatively simple expression for the projector matrix and the tangent stiffness matrix. For CR analyses, this kind of treatment was first introduced by Nour-Omid and Rankin [7], and is well summarized in Felippa and Haugen [38]. We adopt this kind of treatment throughout this work. Assuming that each node has three rotational freedoms, the local rotational displacement $\bar{\boldsymbol{\theta}}_i \in \Re^3$ for a particular node can be obtained through

$$\bar{\boldsymbol{R}}_i = \boldsymbol{R}^{\mathrm{T}} \boldsymbol{R}_i \boldsymbol{R}_0; \quad \bar{\boldsymbol{\Omega}} = \log_{\mathrm{e}}\left(\bar{\boldsymbol{R}}_i\right); \quad \bar{\boldsymbol{\theta}}_i = \operatorname{axial}\left(\bar{\boldsymbol{\Omega}}\right) \tag{14}$$

where $\boldsymbol{R}_i$ represents the current orthogonal rotation matrix associated with the node; $\boldsymbol{R}$ and $\boldsymbol{R}_0$ represent the current and the initial orthogonal rotation matrix of the element, respectively; The operation $\operatorname{axial}(\bullet)$ denotes extracting the axial vector from the spin tensor.

In combination with Eqs. (2) and (14), the relationship between the variation of the local variable ( $\bar{\boldsymbol{v}}$ ) and that of the global variable ( $\boldsymbol{v}$ ) can be derived as $\delta\bar{\boldsymbol{v}} = \bar{\boldsymbol{H}}\bar{\boldsymbol{P}}\operatorname{diag}\left(\boldsymbol{R}\right)^{\mathrm{T}} \delta\boldsymbol{v}$ , where $\delta\bar{\boldsymbol{v}} = \left[\delta\bar{\boldsymbol{u}}_1^{\mathrm{T}}, \delta\bar{\boldsymbol{\theta}}_1^{\mathrm{T}}, \cdots, \delta\bar{\boldsymbol{u}}_N^{\mathrm{T}}, \delta\bar{\boldsymbol{\theta}}_N^{\mathrm{T}}\right]^{\mathrm{T}}$ , $\delta\boldsymbol{v} = \left[\delta\boldsymbol{u}_1^{\mathrm{T}}, \delta\boldsymbol{\omega}_1^{\mathrm{T}}, \cdots, \delta\boldsymbol{u}_N^{\mathrm{T}}, \delta\boldsymbol{\omega}_N^{\mathrm{T}}\right]^{\mathrm{T}}$. As a result, the element internal force is expressed as

$$\boldsymbol{f}\left(\boldsymbol{v}\right) = \operatorname{diag}\left(\boldsymbol{R}\right) \bar{\boldsymbol{P}}^{\mathrm{T}} \bar{\boldsymbol{H}}^{\mathrm{T}} \bar{\boldsymbol{K}} \bar{\boldsymbol{v}} \tag{15}$$

In above expressions, $\bar{\boldsymbol{P}} \in \Re^{6N\times 6N}$ represents the projector matrix. As in the case of continuum elements, it serves as an important quantity to ensure the self-balance of the internal force, and takes the form $\bar{\boldsymbol{P}} = \boldsymbol{I}_{6N} - \bar{\boldsymbol{S}}\bar{\boldsymbol{G}}$ . Here, the moment-arm matrix $\bar{\boldsymbol{S}} \in \Re^{6N\times 3}$ and the spin-fitter matrix $\bar{\boldsymbol{G}} \in \Re^{3\times 6N}$ should be modified to account for the

rotational freedom, as

$$\bar{\boldsymbol{S}} = \left[ \text{spin}\left(\bar{\boldsymbol{x}}_1\right) \quad \boldsymbol{I}_3 \quad \text{spin}\left(\bar{\boldsymbol{x}}_2\right) \quad \boldsymbol{I}_3 \quad \cdots \quad \text{spin}\left(\bar{\boldsymbol{x}}_N\right) \quad \boldsymbol{I}_3 \right]^{\mathrm{T}} \tag{16}$$

$$\bar{\boldsymbol{G}} = \left[ \frac{\partial \bar{\boldsymbol{\omega}}}{\partial \bar{\boldsymbol{u}}_1} \quad \frac{\partial \bar{\boldsymbol{\omega}}}{\partial \bar{\boldsymbol{\theta}}_1} \quad \frac{\partial \bar{\boldsymbol{\omega}}}{\partial \bar{\boldsymbol{u}}_2} \quad \frac{\partial \bar{\boldsymbol{\omega}}}{\partial \bar{\boldsymbol{\theta}}_2} \quad \cdots \quad \frac{\partial \bar{\boldsymbol{\omega}}}{\partial \bar{\boldsymbol{u}}_N} \quad \frac{\partial \bar{\boldsymbol{\omega}}}{\partial \bar{\boldsymbol{\theta}}_N} \right] \tag{17}$$

For shell elements with typical choices of local CR frame, the element rotation is only influenced by the translational displacement of the nodes, and the term $\partial \bar{\boldsymbol{\omega}} / \partial \bar{\boldsymbol{\theta}}_i \left( i = 1, \cdots, N \right)$ in $\bar{\boldsymbol{G}}$ would be zero. This results in an outcome that only the translational force terms are modified for a local force ( $\bar{\boldsymbol{f}}$ ) being pre-multiplied by the transpose of the projector matrix (i.e., $\bar{\boldsymbol{f}} \rightarrow \bar{\boldsymbol{P}}^{\mathrm{T}} \bar{\boldsymbol{f}}$ ).

Different from the case of continuum elements, matrix $\bar{\boldsymbol{H}} \in \Re^{6N \times 6N}$ emerges in the internal force calculation. It links the variation of nodal deformational rotational vector in response to the variation of the element instantaneous spin axial vector at the local level, and is given as

$$\bar{\boldsymbol{H}} = \text{diag}\left[ \boldsymbol{I}_3 \quad \bar{\boldsymbol{H}}_1 \quad \boldsymbol{I}_3 \quad \bar{\boldsymbol{H}}_2 \quad \cdots \quad \boldsymbol{I}_3 \quad \bar{\boldsymbol{H}}_N \right], \quad \bar{\boldsymbol{H}}_i \left( \bar{\boldsymbol{\theta}}_i \right) = \partial \bar{\boldsymbol{\theta}}_i / \partial \bar{\boldsymbol{\omega}} \in \Re^{3 \times 3} \tag{18}$$

For problems with small deformational rotation axial vectors, matrix $\bar{\boldsymbol{H}}$ get close to the identity matrix, and therefore can be neglected in most cases, as is down by many authors. In their review article, Felippa and Haugen ([38], Section 6.3) emphasized the relative importance of using matrix $\bar{\boldsymbol{H}}$ , but without giving a deeper explanation. However, in the authors' opinion, this term should be dropped, not just for simplicity, but for the consistency between the linear and the nonlinear formulations. By "consistency", we mean that the outcomes of the nonlinear CR analysis should be able to deduce to the linear analysis if the element is completely un-rotated. This kind of consistency is met for continuum elements, but is violated for structural elements. This can be justified using the simple example shown in Fig. 5. In this example, all three nodes of the shell element are fixed, thus the element local frame is completely un-rotated. The element is subjected to an external moment $\boldsymbol{M}_{\mathrm{EXT}}$. By dropping the term of matrix $\bar{\boldsymbol{H}}$ , the nonlinear CR analysis would lead to the displacement results being identical with those of the linear analysis, which is, however, not the case for the CR analysis including $\bar{\boldsymbol{H}}$ . Why including matrix $\bar{\boldsymbol{H}}$ rigorously derived from the kinematic relationship should violate the consistency? This question deserves further exploring, and the authors' preliminary explanation is that the inconsistency comes from the fact that the definition of moment [40-42] for the local/linear and nonlinear formulation is inconsistent. In the local/linear formulation, the direction of internal moment is fixed to the local frame, while in the nonlinear formulation with matrix $\bar{\boldsymbol{H}}$ the direction of internal moment is rotation-dependent.

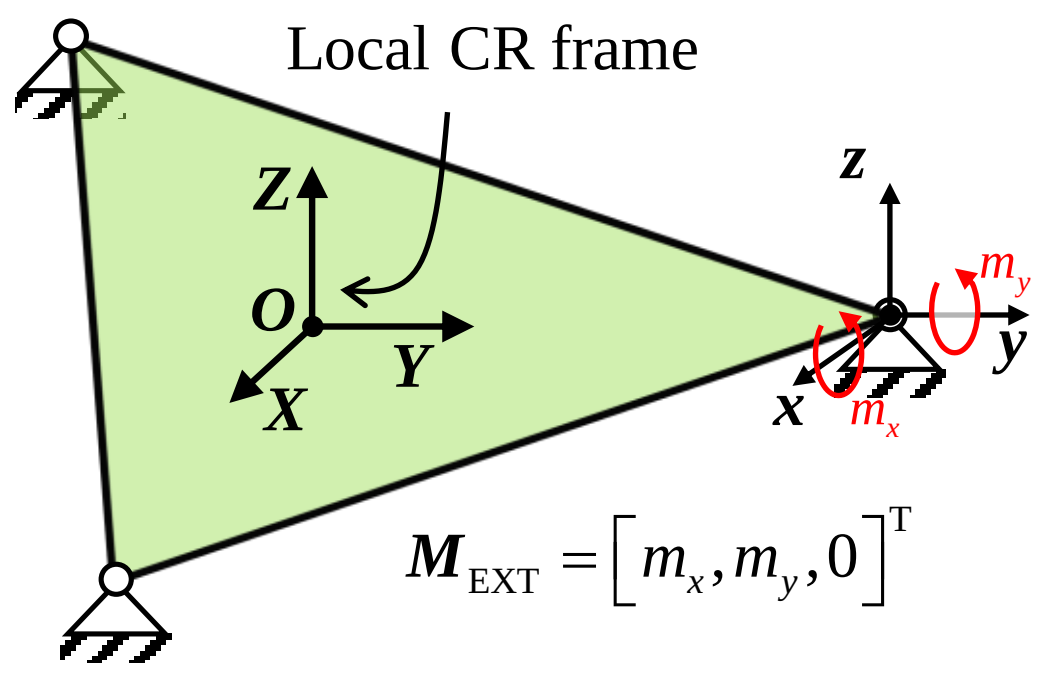

Fig. 5 A shell element subjected to an external moment, with all three nodes being fixed.

The consistent tangent stiffness matrix can be derived by taking the derivative of the internal force (15) with the global variable. This involves a further variation of the projector matrix, which is cumbersome and heavily depends on the choice of the local CR frame. This motivates our following work.

## 3. CR framework based on the direct force correction approach

In the developed direct force correction based CR framework. The projector matrix is first discarded in the internal force calculation. For structural elements, matrix $\bar{\boldsymbol{H}}$ is also dropped based on our previous discussions. The calculation of the element internal force is therefore deduced into

$$\boldsymbol{f} = \mathrm{diag}(\boldsymbol{R})\,\bar{\boldsymbol{f}} = \mathrm{diag}(\boldsymbol{R})\,\bar{\boldsymbol{K}}\bar{\boldsymbol{v}} \tag{19}$$

We remark this formulation exactly coincides with many early works [2, 5] involving CR analysis and is identical with the C formulation summarized in ([38], Section 6.1). This formulation has a straightforward mechanical significance relevant to the core concept of CR approach: the local internal force $\bar{\boldsymbol{f}} = \bar{\boldsymbol{K}}\bar{\boldsymbol{v}}$ physically means the translational force and the moment expressed in the local CR frame, and the global internal force is simply a rotation of these local forces. The performance of using this formulation heavily relies on a "good" choice of the local CR frame. As previously mentioned, other than some particular elements with special choices of local CR frame, this formulation will generally be accompanied with non-balanced moment for the element which is exactly the major factor that destroys the accuracy of this formulation. In CR analysis, the conventional projection matrix $\bar{\boldsymbol{P}}$ works because it essentially makes a correction for the local internal force such that the non-balanced property of the force is eliminated. The drawback of using conventional projection matrix is that this correction is fully coupled with the choice of the local CR frame, leading to rather cumbersome expressions. The main idea of the proposed approach is that one may directly make the correction in the *global* frame and using some simple rules that is independent of the choice of the local CR frame. We focus on the analysis of structural elements, and the degeneration to continuum elements is straightforward by dropping the terms related to the rotational freedoms.

### 3.1 General formulation of the correction force

For easy of presentation, we replace the commonly used global nodal displacements with the global nodal position as the element global variable, i.e.

$$\boldsymbol{x} = \left[ \boldsymbol{x}_1^{\mathrm{T}}, \boldsymbol{\omega}_1^{\mathrm{T}}, \cdots, \boldsymbol{x}_N^{\mathrm{T}}, \boldsymbol{\omega}_N^{\mathrm{T}} \right]^{\mathrm{T}} \tag{20}$$

where $\boldsymbol{x}_i$ denotes the position of the *i*th node. For each current configuration $\boldsymbol{x}$, the global internal force $\boldsymbol{f}$ is first obtained from Eq. (19) and is referred to as the preliminary internal force in the following.

$$\boldsymbol{f} = \begin{bmatrix} \boldsymbol{n}_1^{\mathrm{T}} & \boldsymbol{m}_1^{\mathrm{T}} & \boldsymbol{n}_2^{\mathrm{T}} & \boldsymbol{m}_2^{\mathrm{T}} & \cdots & \boldsymbol{n}_N^{\mathrm{T}} & \boldsymbol{m}_N^{\mathrm{T}} \end{bmatrix}^{\mathrm{T}} \tag{21}$$

where $\boldsymbol{n}_i$ and $\boldsymbol{m}_i$ denote the translational force and the moment of the *i*th node expressed in the global frame. The element self-balance condition defines a constraint ($\boldsymbol{g} \in \Re^6$) in term of the element internal force ($\boldsymbol{f}$) and the element current global variables ($\boldsymbol{x}$):

$$\boldsymbol{g}(\boldsymbol{f}, \boldsymbol{x}) \equiv \begin{bmatrix} \sum_{i=1}^{N} \boldsymbol{n}_i \\ \sum_{i=1}^{N} \left( \boldsymbol{x}_i \times \boldsymbol{n}_i + \boldsymbol{m}_i \right) \end{bmatrix} = \boldsymbol{0} \tag{22}$$

where the first part represents a balance of the translational force, which is always satisfied for the preliminary internal force $\boldsymbol{f}$ calculated from Eq. (19) regardless how the local CR frame is chosen; the second part represents a balance of the element moment, which is typically unsatisfied due to an inappropriate choice of the local CR frame. It is worth noting that $\boldsymbol{g}$ is a linear function of $\boldsymbol{f}$ for a current fixed $\boldsymbol{x}$.

The objective of the correction process is to find a correction term $\underline{\boldsymbol{f}}$ to the preliminary force, such that the final internal force $\boldsymbol{f}_{\mathrm{final}} = \boldsymbol{f} + \underline{\boldsymbol{f}}$ is balanced in terms both of the translational force and the moment. We perform the process using the minimal norm correction strategy, which results in the following optimization problem

$$\begin{cases} \text{find}: \underline{\boldsymbol{f}} \\ \min: \underline{\boldsymbol{f}}^{\mathrm{T}} \boldsymbol{W} \underline{\boldsymbol{f}} / 2 \\ \text{s.t.} \ : \boldsymbol{g}\left( \boldsymbol{f} + \underline{\boldsymbol{f}}, \boldsymbol{x} \right) = \boldsymbol{0} \end{cases} \tag{23}$$

where $\boldsymbol{W} \in \Re^{6N \times 6N}$ is the designated weight matrix which is assumed to be a diagonal matrix. We first focus on the expression with general weight matrix. Discussion and simplification on different choices of weight matrix will be given subsequently. The above optimization problem can be solved by using Lagrangian multiplier method.

$$\begin{cases} \ell = \underline{\boldsymbol{f}}^{\mathrm{T}} \boldsymbol{W} \underline{\boldsymbol{f}} / 2 + \boldsymbol{\lambda}^{\mathrm{T}} \boldsymbol{g}\left( \boldsymbol{f} + \underline{\boldsymbol{f}}, \boldsymbol{x} \right) \\ \dfrac{\partial \ell}{\partial \underline{\boldsymbol{f}}} = \boldsymbol{0}; \dfrac{\partial \ell}{\partial \boldsymbol{\lambda}} = \boldsymbol{0} \end{cases} \tag{24}$$

This results in a linear system in terms of the unknown quantity $\underline{\boldsymbol{f}}$ and $\lambda$:

$$\begin{bmatrix} \boldsymbol{W} & \boldsymbol{g}_{\underline{f}}^{\mathrm{T}} \\ \boldsymbol{g}_{\underline{f}} & \boldsymbol{0} \end{bmatrix} \begin{bmatrix} \underline{\boldsymbol{f}} \\ \lambda \end{bmatrix} + \begin{bmatrix} \boldsymbol{0} \\ \boldsymbol{g}\left( \boldsymbol{f}, \boldsymbol{x} \right) \end{bmatrix} = \begin{bmatrix} \boldsymbol{0} \\ \boldsymbol{0} \end{bmatrix} \tag{25}$$

where $\boldsymbol{g}_{\underline{f}} = \partial \boldsymbol{g} / \partial \underline{\boldsymbol{f}} \in \Re^{6 \times 6N}$ and it is a function of only the current global coordinate $\boldsymbol{x}$.

$$\boldsymbol{g}_{\underline{f}} = \begin{bmatrix} \boldsymbol{I}_3 & \boldsymbol{0}_{3\times3} & \boldsymbol{I}_3 & \boldsymbol{0}_{3\times3} & \cdots & \boldsymbol{I}_3 & \boldsymbol{0}_{3\times3} \\ \mathrm{spin}\left( \boldsymbol{x}_1 \right) & \boldsymbol{I}_3 & \mathrm{spin}\left( \boldsymbol{x}_2 \right) & \boldsymbol{I}_3 & \cdots & \mathrm{spin}\left( \boldsymbol{x}_N \right) & \boldsymbol{I}_3 \end{bmatrix} \tag{26}$$

Noting that $\boldsymbol{g}\left( \boldsymbol{f} \right) \equiv \boldsymbol{g}_{\underline{f}} \boldsymbol{f}$, the solution of problem (25) can be explicitly obtained as

$$\underline{\boldsymbol{f}} = -\boldsymbol{W}^{-1} \boldsymbol{g}_{\underline{f}}^{\mathrm{T}} \left( \boldsymbol{g}_{\underline{f}} \boldsymbol{W}^{-1} \boldsymbol{g}_{\underline{f}}^{\mathrm{T}} \right)^{-1} \boldsymbol{g}_{\underline{f}} \boldsymbol{f} \tag{27}$$

with

$$\lambda = \left( \boldsymbol{g}_{\underline{f}} \boldsymbol{W}^{-1} \boldsymbol{g}_{\underline{f}}^{\mathrm{T}} \right)^{-1} \boldsymbol{g}_{\underline{f}} \boldsymbol{f} \tag{28}$$

### 3.2 The consistent tangent stiffness matrix of the correction force

We now derive the consistent tangent stiffness matrix related to the correction force, i.e., $\underline{\boldsymbol{K}} = \mathrm{d}\underline{\boldsymbol{f}} / \mathrm{d}\boldsymbol{x}$. We assume that the consistent tangent stiffness matrix related to the preliminary force, $\boldsymbol{K} = \mathrm{d}\boldsymbol{f} / \mathrm{d}\boldsymbol{x}$, has been obtained (its detailed expression is referred to [38], Section 6.1). It would be cumbersome if we directly focus on expression (27) to obtain the term. To derive it succinctly and efficiently, we move back to expression (25).

By taking the full derivative of expression (25) with the current global variable $\boldsymbol{x}$ and rearranging, we have

$$\begin{bmatrix} \boldsymbol{W} & \boldsymbol{g}_{\underline{f}}^{\mathrm{T}} \\ \boldsymbol{g}_{\underline{f}} & \boldsymbol{0} \end{bmatrix} \begin{bmatrix} \mathrm{d}\underline{\boldsymbol{f}} / \mathrm{d}\boldsymbol{x} \\ \mathrm{d}\lambda / \mathrm{d}\boldsymbol{x} \end{bmatrix} + \begin{bmatrix} \boldsymbol{M}_1 \\ \boldsymbol{M}_2 \end{bmatrix} = \begin{bmatrix} \boldsymbol{0} \\ \boldsymbol{0} \end{bmatrix} \tag{29}$$

where

$$\boldsymbol{M}_1 = \left. \frac{\partial \left( \boldsymbol{g}_{\underline{f}}^{\mathrm{T}} \boldsymbol{\lambda} \right)}{\partial \boldsymbol{x}} \right|_{\lambda} = \mathrm{diag}\begin{bmatrix} \mathrm{spin}\left( \boldsymbol{\lambda}_{(4:6)} \right) & \boldsymbol{0}_{3\times3} & \cdots & \mathrm{spin}\left( \boldsymbol{\lambda}_{(4:6)} \right) & \boldsymbol{0}_{3\times3} \end{bmatrix} \tag{30}$$

and

$$\boldsymbol{M}_2 = \frac{\mathrm{d}\boldsymbol{g}(\boldsymbol{f})}{\mathrm{d}\boldsymbol{x}} + \left.\frac{\partial\left(\boldsymbol{g}_{\underline{f}}\underline{\boldsymbol{f}}\right)}{\partial\boldsymbol{x}}\right|_{\underline{f}} = \frac{\partial\boldsymbol{g}}{\partial\boldsymbol{f}}\frac{\partial\boldsymbol{f}}{\partial\boldsymbol{x}} + \frac{\partial\boldsymbol{g}(\boldsymbol{f})}{\partial\boldsymbol{x}} + \left.\frac{\partial\left(\boldsymbol{g}_{\underline{f}}\underline{\boldsymbol{f}}\right)}{\partial\boldsymbol{x}}\right|_{\underline{f}} = \boldsymbol{g}_{\underline{f}}\boldsymbol{K} + \boldsymbol{g}_{x} \tag{31}$$

with the relationship $\partial\boldsymbol{g}/\partial\boldsymbol{f} = \boldsymbol{g}_f = \boldsymbol{g}_{\underline{f}}$ being used. In Eq. (30), $\boldsymbol{\lambda}_{(4:6)}$ denotes the $3\times1$ vector construed by the 4~6th component of $\boldsymbol{\lambda}$. By noting that $\boldsymbol{g}_{\underline{f}}\underline{\boldsymbol{f}} = \boldsymbol{g}\left(\underline{\boldsymbol{f}}\right)$, in Eq. (31) $\boldsymbol{g}_x$ is defined as $\partial\boldsymbol{g}\left(\boldsymbol{f}+\underline{\boldsymbol{f}}\right)/\partial\boldsymbol{x}$, and is formed as

$$\boldsymbol{g}_x = \begin{bmatrix} \boldsymbol{0}_{3\times3} & \boldsymbol{0}_{3\times3} & \boldsymbol{0}_{3\times3} & \boldsymbol{0}_{3\times3} & \cdots & \boldsymbol{0}_{3\times3} & \boldsymbol{0}_{3\times3} \\ -\mathrm{spin}\left(\boldsymbol{n}_1+\underline{\boldsymbol{n}}_1\right) & \boldsymbol{0}_{3\times3} & -\mathrm{spin}\left(\boldsymbol{n}_2+\underline{\boldsymbol{n}}_2\right) & \boldsymbol{0}_{3\times3} & \cdots & -\mathrm{spin}\left(\boldsymbol{n}_N+\underline{\boldsymbol{n}}_N\right) & \boldsymbol{0}_{3\times3} \end{bmatrix} \tag{32}$$

where $\underline{\boldsymbol{n}}_i \in \Re^3$ is defined as the translational force of the *i*th node extracted from the correction force $\underline{\boldsymbol{f}}$ ( analogy to Eq. (21)).

By solving Eq. (29), the consistent tangent stiffness matrix related to the correction force can be obtained as

$$\underline{\boldsymbol{K}} = \frac{\mathrm{d}\underline{\boldsymbol{f}}}{\mathrm{d}\boldsymbol{x}} = -\boldsymbol{W}^{-1}\left[\left(\boldsymbol{I}_{6N} - \boldsymbol{g}_{\underline{f}}^{\mathrm{T}}\left(\boldsymbol{g}_{\underline{f}}\boldsymbol{W}^{-1}\boldsymbol{g}_{\underline{f}}^{\mathrm{T}}\right)^{-1}\boldsymbol{g}_{\underline{f}}\boldsymbol{W}^{-1}\right)\boldsymbol{M}_1 + \boldsymbol{g}_{\underline{f}}^{\mathrm{T}}\left(\boldsymbol{g}_{\underline{f}}\boldsymbol{W}^{-1}\boldsymbol{g}_{\underline{f}}^{\mathrm{T}}\right)^{-1}\boldsymbol{M}_2\right] \tag{33}$$

It is worthy pointing that in general both $\boldsymbol{K}$ and $\underline{\boldsymbol{K}}$ are non-symmetric, thus the final consistent tangent stiffness matrix of the element $\boldsymbol{K}_{\mathrm{final}} = \boldsymbol{K} + \underline{\boldsymbol{K}}$ is non-symmetric. Different from the using of conventional nonlinear projection matrix, using the symmetric part of the tangent matrix cannot maintain quadratic convergence. This may be considered as the price to pay for using the present unified force correction approach.

The calculation of the correction force and its tangent stiffness matrix involves an inversion of a $6\times6$ symmetric matrix $\boldsymbol{g}_{\underline{f}}\boldsymbol{W}^{-1}\boldsymbol{g}_{\underline{f}}^{\mathrm{T}}$. To possibly make computations efficient, we may exploit the special form of this matrix as

$$\boldsymbol{g}_{\underline{f}}\boldsymbol{W}^{-1}\boldsymbol{g}_{\underline{f}}^{\mathrm{T}} = \begin{bmatrix} \boldsymbol{D} & \boldsymbol{B} \\ \boldsymbol{B}^{\mathrm{T}} & \boldsymbol{A} \end{bmatrix} \tag{34}$$

where $\boldsymbol{D} \in \Re^{3\times3}$ is a diagonal matrix. The inversion can be obtained as

$$\left(\boldsymbol{g}_{\underline{f}}\boldsymbol{W}^{-1}\boldsymbol{g}_{\underline{f}}^{\mathrm{T}}\right)^{-1} = \begin{bmatrix} \boldsymbol{D}^{-1}\left(\boldsymbol{I}_3 + \boldsymbol{B}\boldsymbol{H}\boldsymbol{B}^{\mathrm{T}}\boldsymbol{D}^{-1}\right) & -\boldsymbol{D}^{-1}\boldsymbol{B}\boldsymbol{H} \\ -\left(\boldsymbol{D}^{-1}\boldsymbol{B}\boldsymbol{H}\right)^{\mathrm{T}} & \boldsymbol{H} \end{bmatrix} \tag{35}$$

with

$$\boldsymbol{H} = \left(\boldsymbol{A} - \boldsymbol{B}^{\mathrm{T}}\boldsymbol{D}^{-1}\boldsymbol{B}\right)^{-1} \tag{36}$$

### 3.3 Different choices of the weight matrix

### 3.3.1 Case I

The most direct way of choosing weight matrix is setting $\boldsymbol{W} = \boldsymbol{I}_{6N}$ . In this situation, both the translational force term and the moment term of the preliminary force are corrected with an equal weight. This results in the final element internal force being

$$\boldsymbol{f}_{\mathrm{final}} = \boldsymbol{f} + \underline{\boldsymbol{f}} = \left(\boldsymbol{I}_{6N} - \boldsymbol{g}_{\underline{f}}^{\mathrm{T}}\left(\boldsymbol{g}_{\underline{f}}\boldsymbol{g}_{\underline{f}}^{\mathrm{T}}\right)^{-1}\boldsymbol{g}_{\underline{f}}\right)\boldsymbol{f} \triangleq \boldsymbol{P}_{\mathrm{linear}}^{\mathrm{T}}\boldsymbol{f} \tag{37}$$

In [6], Rankin and Nour-Omid also developed a linear projector matrix by orthogonalizing of the displacement against the rigid body modes. It is interesting to find that the formulation of the linear projector matrix therein shows a similarity with of the proposed direct force approach if we interpret the term $\boldsymbol{g}_{\underline{f}}$ as the set of rigid body modes ( $\boldsymbol{\Gamma}$ ) therein. However, they only restrict the application of the linear projector matrix to analyses with infinitesimal rigid body rotation, and the possible potentiality of the formulation in finite rotation analyses is not explored.

For the present choice of weight matrix, the final element consistent tangent stiffness matrix can be given as

$$\begin{aligned}\boldsymbol{K}_{\mathrm{final}} = \boldsymbol{K} + \underline{\boldsymbol{K}} &= \left(\boldsymbol{I}_{6N} - \boldsymbol{g}_{\underline{f}}^{\mathrm{T}}\left(\boldsymbol{g}_{\underline{f}}\boldsymbol{g}_{\underline{f}}^{\mathrm{T}}\right)^{-1}\boldsymbol{g}_{\underline{f}}\right)\boldsymbol{K} \\ &-\left(\boldsymbol{I}_{6N} - \boldsymbol{g}_{\underline{f}}^{\mathrm{T}}\left(\boldsymbol{g}_{\underline{f}}\boldsymbol{g}_{\underline{f}}^{\mathrm{T}}\right)^{-1}\boldsymbol{g}_{\underline{f}}\right)\boldsymbol{M}_1 - \boldsymbol{g}_{\underline{f}}^{\mathrm{T}}\left(\boldsymbol{g}_{\underline{f}}\boldsymbol{g}_{\underline{f}}^{\mathrm{T}}\right)^{-1}\boldsymbol{g}_{x}\end{aligned} \tag{38}$$

where the first term of the right hand of the second equal sign stems from variation of the preliminary internal force ( $\boldsymbol{f}$ ) in Eq. (37) and the rest two terms stem from the variation of the linear projector matrix ( $\boldsymbol{P}_{\mathrm{linear}}$ ) in Eq. (37) .

### 3.3.2 Case II

Another kind of approach for structural elements is only making the correction on those moment terms with an equal weight, which corresponds to the following inversion of the weight matrix

$$\boldsymbol{W}^{-1} = \mathrm{diag}\left[\boldsymbol{W}_1^{-1} \quad \boldsymbol{W}_2^{-1} \quad \cdots \quad \boldsymbol{W}_N^{-1}\right] \tag{39}$$

with the submatrix being

$$\boldsymbol{W}_i^{-1} = \begin{bmatrix} \boldsymbol{0}_{3\times3} & \boldsymbol{0}_{3\times3} \\ \boldsymbol{0}_{3\times3} & \boldsymbol{I}_3 \end{bmatrix} \in \Re^{6\times6}, \quad \left(i = 1, \cdots, N\right) \tag{40}$$

This results in an extremely simple expression of the correction force

$$\underline{\boldsymbol{f}} = \left[ \boldsymbol{0}_{1\times3} \quad -\boldsymbol{m}_{\text{unblance}}^{\text{T}} \quad \boldsymbol{0}_{1\times3} \quad -\boldsymbol{m}_{\text{unblance}}^{\text{T}} \quad \cdots \quad \boldsymbol{0}_{1\times3} \quad -\boldsymbol{m}_{\text{unblance}}^{\text{T}} \right]^{\text{T}} \Big/ N \tag{41}$$

where $\boldsymbol{m}_{\text{unblance}} \in \Re^3$ is the simply the unbalanced moment of the element, i.e., $\sum_{i=1}^{N} \left( \boldsymbol{x}_i \times \boldsymbol{n}_i + \boldsymbol{m}_i \right)$. The degenerative expression of the consistent tangent stiffness matrix of the correction force is given as

$$\underline{\boldsymbol{K}} = \left( -\boldsymbol{\Gamma}\boldsymbol{K} + \boldsymbol{\Psi} \right) \big/ N \tag{42}$$

where

$$\boldsymbol{\Gamma} = \begin{bmatrix} \boldsymbol{\Gamma}_1 \\ \vdots \\ \boldsymbol{\Gamma}_N \end{bmatrix} \in \Re^{6N\times6N} \quad \text{and} \quad \boldsymbol{\Psi} = \begin{bmatrix} \boldsymbol{\Psi}_1 \\ \vdots \\ \boldsymbol{\Psi}_N \end{bmatrix} \in \Re^{6N\times6N} \tag{43}$$

with

$$\boldsymbol{\Gamma}_i = \begin{bmatrix} \boldsymbol{0}_{3\times3} & \boldsymbol{0}_{3\times3} & \boldsymbol{0}_{3\times3} & \boldsymbol{0}_{3\times3} & \cdots & \boldsymbol{0}_{3\times3} & \boldsymbol{0}_{3\times3} \\ \text{spin}(\boldsymbol{x}_1) & \boldsymbol{I}_3 & \text{spin}(\boldsymbol{x}_2) & \boldsymbol{I}_3 & \cdots & \text{spin}(\boldsymbol{x}_N) & \boldsymbol{I}_3 \end{bmatrix} \tag{44}$$

$$\boldsymbol{\Psi}_i = \begin{bmatrix} \boldsymbol{0}_{3\times3} & \boldsymbol{0}_{3\times3} & \boldsymbol{0}_{3\times3} & \boldsymbol{0}_{3\times3} & \cdots & \boldsymbol{0}_{3\times3} & \boldsymbol{0}_{3\times3} \\ \text{spin}(\boldsymbol{n}_1) & \boldsymbol{0}_{3\times3} & \text{spin}(\boldsymbol{n}_2) & \boldsymbol{0}_{3\times3} & \cdots & \text{spin}(\boldsymbol{n}_N) & \boldsymbol{0}_{3\times3} \end{bmatrix} \tag{45}$$

The prominent advantage of using this kind of correction approach for structural elements comes from its computational efficiency. Almost no computational effort is gained for the correction process. This is in particular for explicit dynamic analyses, where only the correction of internal forces is involved.

### 3.3.3 Case III

The force correction approach with previous weight matrices works well for general continuum elements and structural elements. However, problems may occur for analyses of some shell elements with drilling rotations. If the deformation of the structure is limited in a plane and the element is subjected to only membrane strain, the moment terms related to the drilling degree of freedoms may be over-corrected, which may lead to convergence difficulty for the global Newton-Raphson iteration. In this case, one may only make corrections on those translational force terms with an equal weight, which corresponds to the sub matrix in Eq. (39) being

$$\boldsymbol{W}_i^{-1} = \begin{bmatrix} \boldsymbol{I}_3 & \boldsymbol{0}_{3\times3} \\ \boldsymbol{0}_{3\times3} & \boldsymbol{0}_{3\times3} \end{bmatrix} \in \Re^{6\times6}, \quad \left( i = 1, \cdots, N \right) \tag{46}$$

For shell elements, this kind of correction approach shows some similarities with the use of conventional nonlinear projection matrix, since both approaches only make corrections on those translational forces. The difference is that for the use of conventional nonlinear projection matrix, the weight matrix implicated in the formulation is constructed based on the choice of the local CR frame, and usually a non-equal weight is used in the correction (only those translational forces of which the degree of freedoms being related to the calculation of the local CR frame are corrected).

It is worthy pointing that a problem of this kind of correction approach occurs for some special elements (e.g., two-node bar, beam or shaft element) where the term $\boldsymbol{g}_{\underline{f}}\boldsymbol{W}^{-1}\boldsymbol{g}_{\underline{f}}^{\mathrm{T}}$ in the correction force calculation will become singular. This problem may be overcome by adopting the Moore-Penrose pseudoinverse in the calculation, which is, however, time-consuming and is not recommended in real applications. In our following numerical examples involving beams elements, this kind of correction approach is also considered for purely investigating the accuracy of the numerical results.

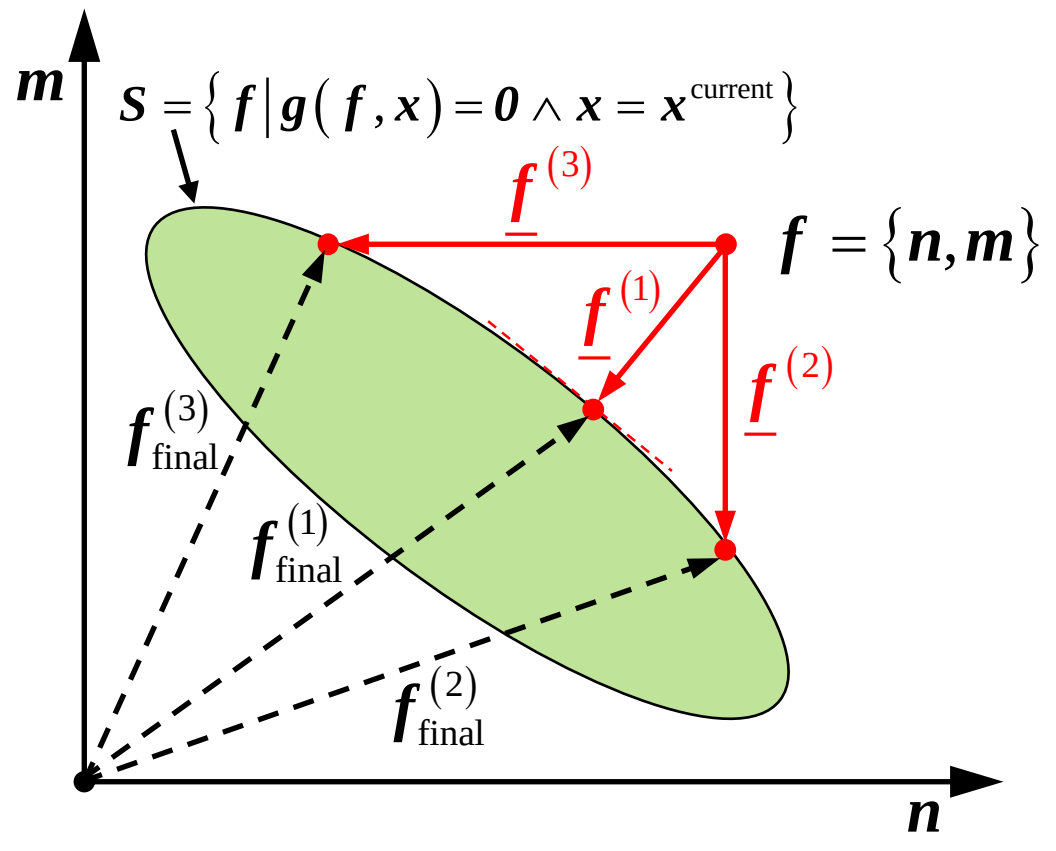

Fig. 6 Geometrical interpretation of the correction approach with the previous considered three different weight matrices.

The previous three correction cases can be geometrically interpreted with Fig. 6. **Table 1** summarizes the direct force correction based algorithm for general CR analysis.

**Table 1** Element independent, direct force correction based algorithm for general CR analysis (with general weight matrix).

1. For each element perform the following:

(a) Calculate $\boldsymbol{R}$ based on the current configuration using some direct or indirect rules; Calculate local displacement $\bar{\boldsymbol{v}}$ .

(b) Evaluate the preliminary internal force ( $\boldsymbol{f}$ ) and its tangent stiffness matrix ( $\boldsymbol{K}$ ).

(c) Form matrix $\boldsymbol{g}_{\underline{f}}$ ; Calculate the Lagrangian multiplier ( $\lambda$ ) and the correction force ( $\underline{\boldsymbol{f}}$ ) .

(d) Form matrix $\boldsymbol{g}_{x}$ , $\boldsymbol{M}_1$ and $\boldsymbol{M}_2$ ; Calculate the tangent stiffness matrix of the correction force ( $\underline{\boldsymbol{K}}$ ) .

(e) Evaluate the final element internal force and tangent stiffness matrix:

$$\boldsymbol{f}_{\text{final}} = \boldsymbol{f} + \underline{\boldsymbol{f}} \,;\quad \boldsymbol{K}_{\text{final}} = \boldsymbol{K} + \underline{\boldsymbol{K}}$$

2. Assemble and solve:

$$\boldsymbol{K}^{\text{total}}\Delta\boldsymbol{v}^{\text{total}} = \boldsymbol{f}^{\text{total}}$$

3. Update system global variable. For a particular node:

$$\boldsymbol{u}_i := \boldsymbol{u}_i + \Delta\boldsymbol{u}_i ; \boldsymbol{R}_i := \exp\left(\text{spin}\left(\Delta\boldsymbol{\omega}\right)\right)\boldsymbol{R}_i$$

4. Terminate if converged

## 4. Numerical examples

Multiple examples involving various classical elements and various choices of local CR frame are presented to demonstrate the performance of the proposed framework. Generally, a moderately coarse mesh of FEM models is used. In this situation the numerical results are sensitive with the choice of the element local CR frame. Limited to the research scope of this paper, we do not make extensive comparisons with other existing works, and emphasis is placed on comparing the numerical results obtained with the following three kinds of method for the internal force calculation under the same FEM model.

(1) Method “S”: The internal force is calculated solely with the “simplest” expression (Eq. (19)), i.e., using the “main part” of the internal force and without using the nonlinear projection matrix or the proposed correction approach.

(2) Method “S+P”: The internal force is calculated using expressions in conjunction with the nonlinear projection matrix. The nonlinear projection matrix must correspond to the choice of the local CR frame.

(3) Method “S+C1/2/3”: The internal force is calculated using expressions in conjunction with the proposed correction approach using different weight matrices, i.e., case 1/2/3.

Besides, for all the examples the final deformation configuration obtained by commercial software ANSYS is presented. Obviously, due to the element formulations being different, we cannot expect an exact agreement between results by proposed approaches and by ANSYS. The small relative errors, however, is able to demonstrate the correctness of the present works. In some cases, the nonlinear projection matrix and its variation are rather complicated to derive, and we use the high-accuracy complex step based finite difference method [43-45](see in the Appendix) to obtain this term. For all the examples, the parameters are considered as dimensionless; the analyses are conducted using multi step loading strategy, and the convergence criterion is defined as $\left\|\boldsymbol{F}_{\text{RES}}\right\| \le \varepsilon$, where $\boldsymbol{F}_{\text{RES}}$ denotes the residual global force vector and $\varepsilon$ denotes the convergence tolerance and is set as $1\times10^{-5}$.

The core linear elements involved in this work are shown in Fig. 7, and the corresponding choices of element local CR frame are summarized below.

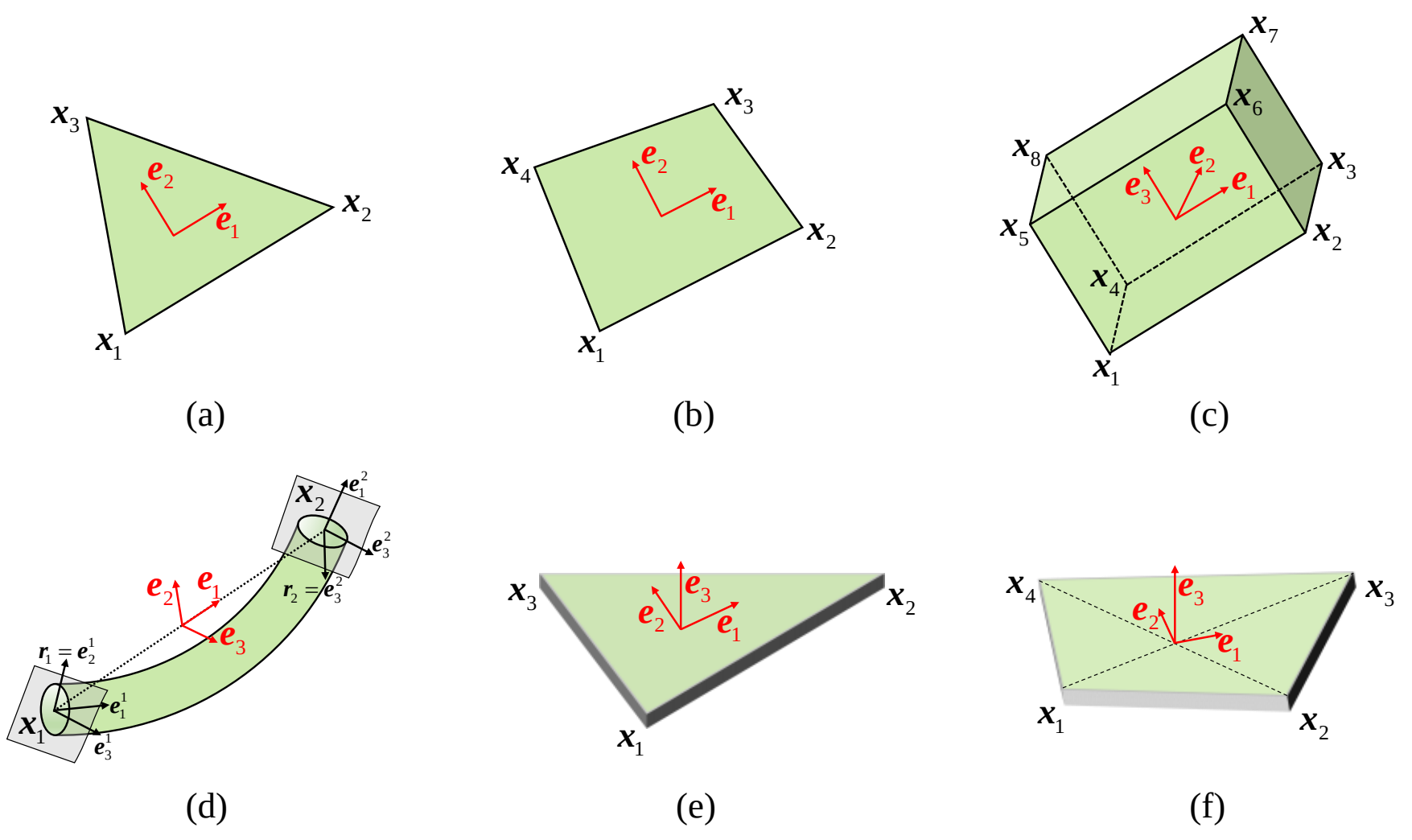

Fig. 7 The elements involved in the numerical examples: (a) 3-node plane triangle element; (b) 4-node plane quadrilateral element; (c) 8-node spatial solid element; (d) 2-node spatial beam element; (e) 3-node spatial triangle shell element; (f) 4-node spatial quadrilateral shell element.

(a). 3-node plane triangle element.

The core linear element is the commonly used constant strain triangle element. The following three different choices of element local CR frame are considered.

(1). Side alignment approach: the element local CR frame is determined by using the side alignment, where the rotation of the element is reflected through the rotation of a particular side. Taking the side 1-2 as an example (which is adopted throughout this work), the current rotational matrix $\boldsymbol{R} \in \Re^{2\times 2}$ are calculated, respectively, as

$$\boldsymbol{R} = \left[\boldsymbol{e}_1, \boldsymbol{e}_2\right] = \left[\frac{\boldsymbol{x}_2 - \boldsymbol{x}_1}{\left\|\boldsymbol{x}_2 - \boldsymbol{x}_1\right\|}, \frac{\boldsymbol{R}_{90^\circ}\left(\boldsymbol{x}_2 - \boldsymbol{x}_1\right)}{\left\|\boldsymbol{x}_2 - \boldsymbol{x}_1\right\|}\right] \tag{47}$$

with $\boldsymbol{R}_{90^\circ} = \begin{bmatrix} 0 & -1 \\ 1 & 0 \end{bmatrix}$. The element initial rotational matrix is determined by an analogous expression but with the initial nodal coordinates.

(2). Least square approach: the element local CR frame is determined by a least square fit of the local deformation. The initial rotational matrix and current rotational matrix can be expressed as

$$\boldsymbol{R}_0 = \boldsymbol{I}_2; \boldsymbol{R} = \arg\min\left(\bar{\boldsymbol{u}}^{\mathrm{T}}\bar{\boldsymbol{u}}\right) \tag{48}$$

where $\bar{\boldsymbol{u}} \in \Re^{6\times 1}$ is the local displacement vector that is compatible with the chosen $\boldsymbol{R}_0$.

(3). Polar decomposition approach: the element local CR frame is determined by using the polar decomposition of the deformation gradient ($\boldsymbol{F} \in \Re^{2\times 2}$) of the element central point.

$$\boldsymbol{R}_0 = \boldsymbol{I}_2; \boldsymbol{R} = \{\text{solve}: \boldsymbol{F} = \boldsymbol{R}\boldsymbol{U}\} \tag{49}$$

where $\boldsymbol{U} \in \Re^{2\times 2}$ denotes the right stretch tensor.

(b). 4-node plane quadrilateral element.

The core linear element is the commonly used bi-linear quadrilateral plane element with full integration. The three choices of element local CR frame are the same as those of the previous 3-node triangle element. Battini [8] presented detailed expressions for CR analysis of this kind of element with the local CR frame being calculated using the least square approach.

(c). 8-node spatial solid element.

The core linear element is the tri-linear hexahedral block element with full integration. We consider two different choices of element local CR frame: the side alignment approach and the polar decomposition approach. For the side alignment approach at three-dimensional case, arbitrarily (non-collinear) three nodes can be used. Taking the nodes 1-2-3 as an example (which is adopted throughout this work), the current element rotational matrix $\boldsymbol{R} = [\boldsymbol{e}_1, \boldsymbol{e}_2, \boldsymbol{e}_3] \in \Re^{3\times 3}$ can be determined with

$$\boldsymbol{e}_1 = \frac{\boldsymbol{x}_2 - \boldsymbol{x}_1}{\|\boldsymbol{x}_2 - \boldsymbol{x}_1\|}; \boldsymbol{e}_3 = \frac{(\boldsymbol{x}_2 - \boldsymbol{x}_1)\times(\boldsymbol{x}_3 - \boldsymbol{x}_1)}{\|(\boldsymbol{x}_2 - \boldsymbol{x}_1)\times(\boldsymbol{x}_3 - \boldsymbol{x}_1)\|}; \boldsymbol{e}_2 = \boldsymbol{e}_3 \times \boldsymbol{e}_1 \tag{50}$$

The element initial rotational matrix is determined by an analogous expression but with the initial nodal coordinates.

(d). 2-node spatial beam element.

The core linear element is the commonly used spatial Euler-Bernoulli beam. We consider one choice of element local CR frame that adopted in [17]. The current element rotational matrix $\boldsymbol{R} = [\boldsymbol{e}_1, \boldsymbol{e}_2, \boldsymbol{e}_3] \in \Re^{3\times 3}$ is determined using

$$\boldsymbol{e}_1 = \frac{\boldsymbol{x}_2 - \boldsymbol{x}_1}{\|\boldsymbol{x}_2 - \boldsymbol{x}_1\|}; \boldsymbol{e}_3 = \frac{\boldsymbol{e}_1 \times \boldsymbol{r}}{\|\boldsymbol{e}_1 \times \boldsymbol{r}\|}; \boldsymbol{e}_2 = \boldsymbol{e}_3 \times \boldsymbol{e}_1 \tag{51}$$

where the auxiliary vector $\boldsymbol{r}$ is calculated as

$$\boldsymbol{r} = \frac{1}{2}(\boldsymbol{r}_1 + \boldsymbol{r}_2); \boldsymbol{r}_i = \boldsymbol{R}_i \boldsymbol{R}_0 [0,1,0]^{\mathrm{T}} \tag{52}$$

The element initial rotational matrix is determined by an analogous expression but with the initial nodal coordinates.

(e). 3-node spatial triangle shell element.

The core linear element is the flat shell element which is a combination of discrete Kirchhoff triangle (DKT) plate and constant strain triangular membrane. We consider one choice of element local CR frame: the side alignment approach. The calculation of the current and initial element rotational matrix is the same as that of the 8-node solid element.

(f). 4-node spatial quadrilateral shell element.

The core linear element is the flat shell element which is a combination of discrete Kirchhoff quadrilateral (DKQ) plate and bi-linear quadrilateral membrane with full integration. We consider one choice of element local CR frame: the side alignment approach that is adopted in [6, 7]. The current element rotational matrix $\boldsymbol{R}=[\boldsymbol{e}_1,\boldsymbol{e}_2,\boldsymbol{e}_3]\in\Re^{3\times3}$ is calculated with

$$\boldsymbol{e}_3=\frac{(\boldsymbol{x}_3-\boldsymbol{x}_1)\times(\boldsymbol{x}_4-\boldsymbol{x}_2)}{\|(\boldsymbol{x}_3-\boldsymbol{x}_1)\times(\boldsymbol{x}_4-\boldsymbol{x}_2)\|};\boldsymbol{e}_1=\frac{(\boldsymbol{x}_4-\boldsymbol{x}_1)\times\boldsymbol{e}_3}{\|(\boldsymbol{x}_4-\boldsymbol{x}_1)\times\boldsymbol{e}_3\|};\boldsymbol{e}_2=\boldsymbol{e}_3\times\boldsymbol{e}_1 \tag{53}$$

The element initial rotational matrix is determined by an analogous expression but with the initial nodal coordinates.

### 4.1. Plane angle frame

We test the performance of the developed framework for plane elements using the example involving a plane angle frame. Similar example was investigated in [8]. As shown in Fig. 8, the structure is clamped at one end and loaded by a uniformly distributed horizontal force at the other end. The maximum force magnitude is taken as $4\times10^4$. The results are obtained using 20 equal loading steps.

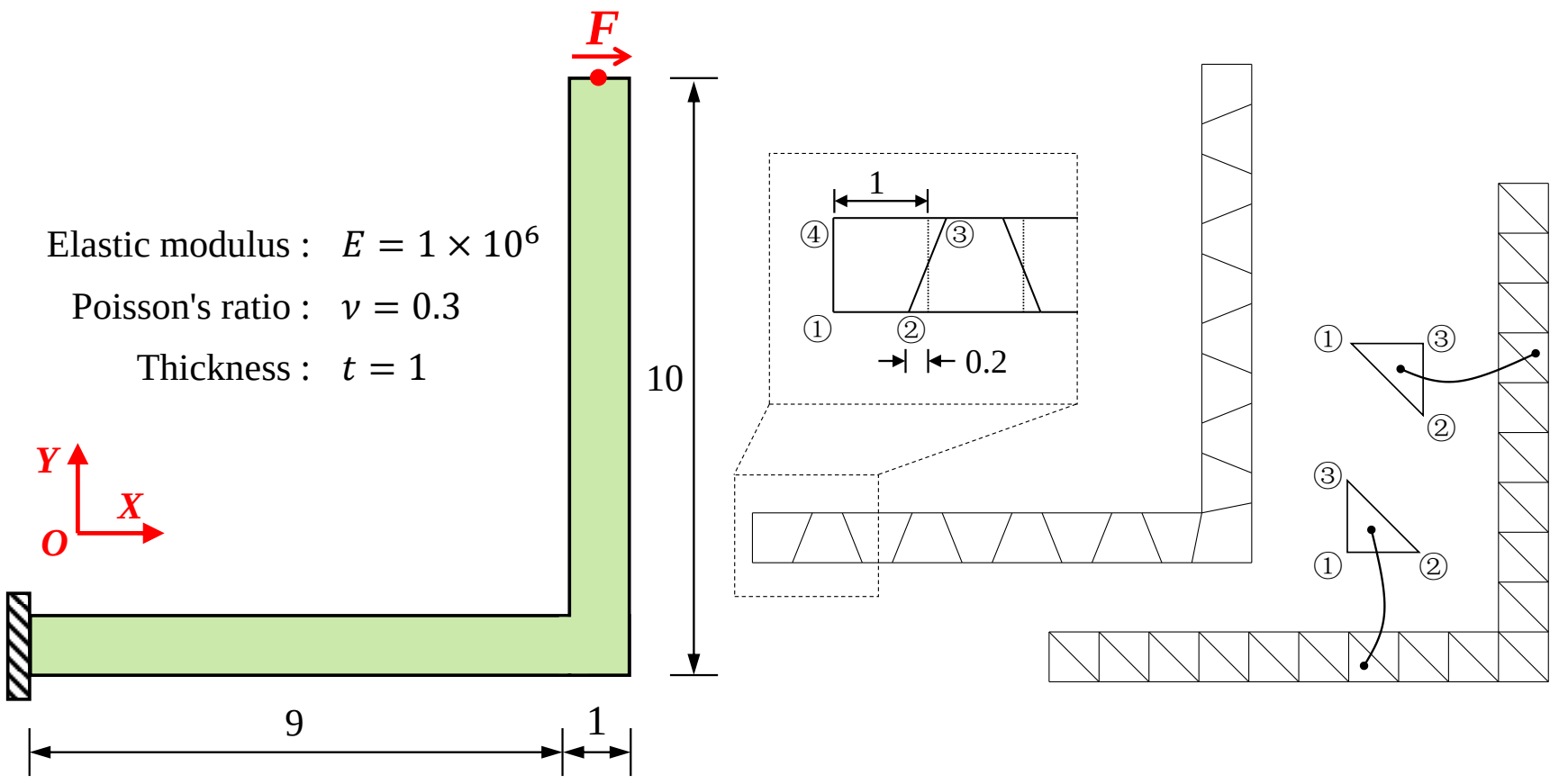

Fig. 8 Plane angle frame: geometrical and material parameters, and FEM models.

We first conduct the analysis using plane triangle elements with the previous mentioned three choices of the local CR frame. For each choice of the local CR frame, three internal force calculation methods, S, S+P, S+C1, are used. First, the results show that for the local CR frame calculated with polar decomposition approach, the three different internal force calculation methods yield *exactly* the same results. This outcome verifies our previous conclusion that for the constant strain plane triangle element the polar decomposition approach can find the completely accurate element rigid-body rotation. Since no unbalance element moment would be produced even use the simplest force expression (19), these is no need to use the proposed force correction approach or the nonlinear projection matrix. Results obtained with this correct rigid-body rotation possesses the best accuracy that the CR formulation can achieve, and it is referred to as the "reference solution" to verify the results obtained with the other choices of local CR frame. Fig. 9 presents the evolution curves of the vertical displacement of the loading point for the other two choices of local CR frame. Fig. 10 plots the final deformed configurations for different cases. To highlight the correctness of the results, the deformed configuration obtained by ANSYS (using plane 42 elements with KEYOPT(2)=1) is also presented. From these results, it can be found that for the side alignment approach the results obtained with S+C1 and S+P methods are close to the reference solution, while the results obtained with S method is far apart with the other results. For the least square approach, the three different force calculation methods yield similar results. This outcome shows that, at least for this particular example, the least square approach can yield a considerably accurate rigid body rotation evaluation. Even so, from the enlarged view of Fig. 9 (b), one can find that the S+C1 and S+P methods yield almost identical results. These results, compared with the result of S method, are more close to the reference solution. The above outcomes show that the effects of using the proposed force correction approach is remarkable, and the performance is comparable with that using conventional nonlinear projection matrix.

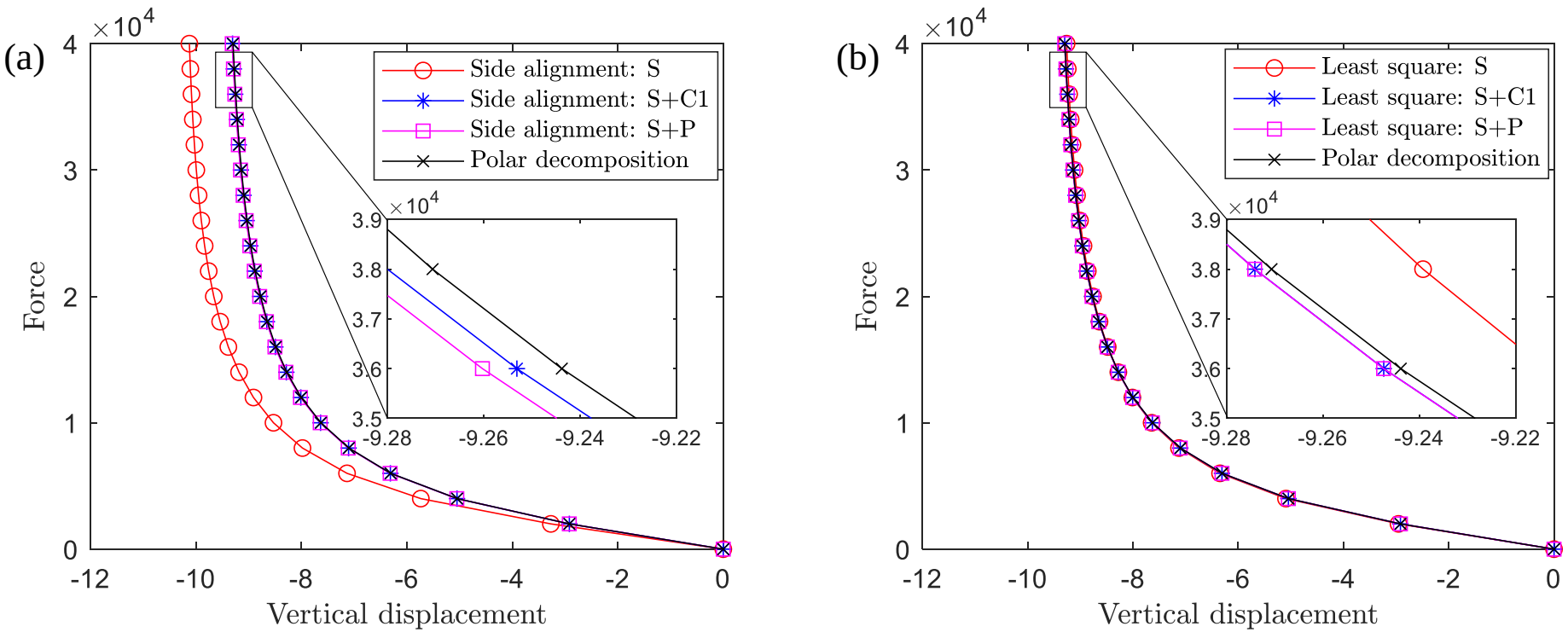

Fig. 9 Evolution curves of the vertical displacement of the loading point with two choices of the local CR frame: (a) side alignment approach; (b) least square approach.

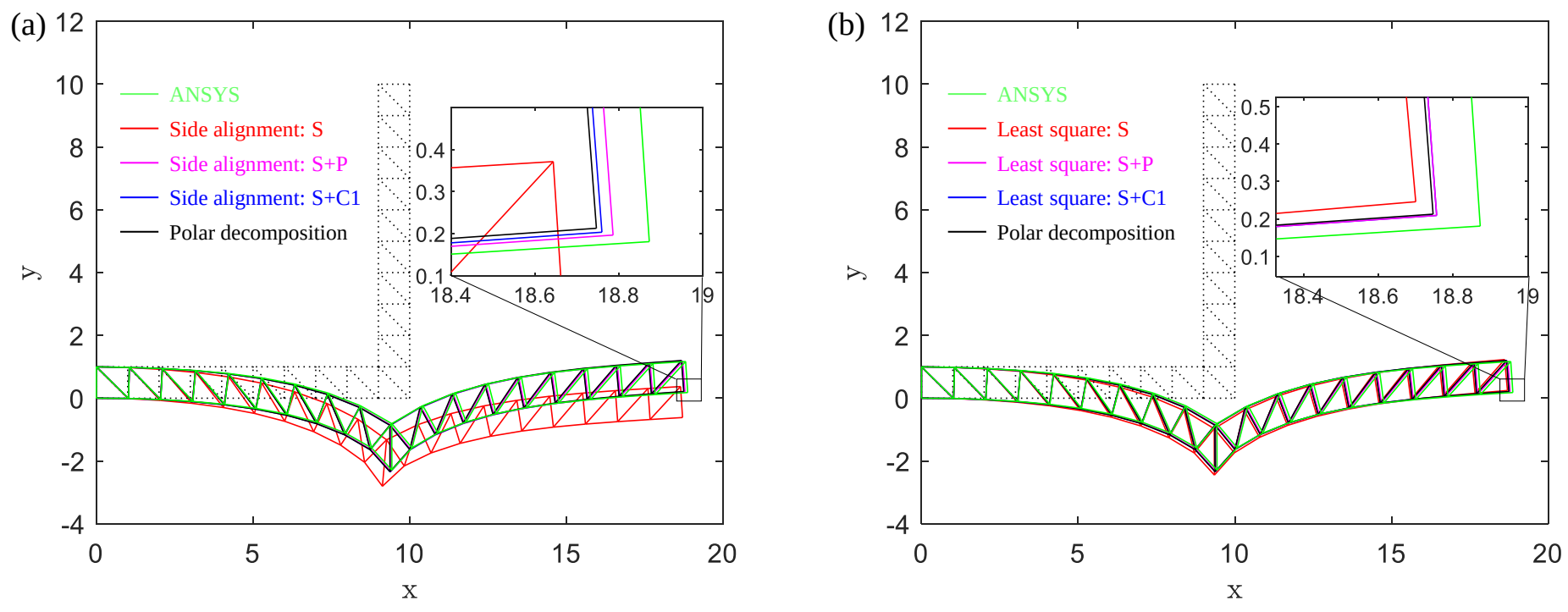

Fig. 10 Deformed configurations at the maximum load: (a) side alignment approach; (b) least square approach.

We next conduct the analysis using plane quadrilateral elements. Firstly, Fig. 11(a) gives the evolution curves of the vertical displacement of the loading point for the polar decomposition approach. Different from the case of using plane triangle elements, the solutions obtained by three different internal force calculation methods are not exactly the same, although they are quite close. This is due to the bi-linear function being used as the element interpolation function, the rotation of each point at the element is generally not the same; the element rotation obtained by the polar decomposition approach actually reflects the rotation of the element central point which in general is not strictly equal to the “mean” rotation of the whole element. Even so, the three results are fairly close, implying that the polar decomposition approach is still a “good” choice for local CR frame determination for this particular example. To verify the other two choices of local CR frame, we use the result obtained by the polar decomposition approach in conjunction with the S+P method as the “reference solution”. Fig. 11(b) and (c) present the evolution curves of the vertical displacement of the loading point, with comparison with the “reference solution”. As can be seen, for both cases the solutions with the proposed force correction process will, in varying degrees, approach to the “reference solution”. The above outcomes also imply that, from the viewpoint of accuracy, the slide alignment approach is the worst choice for the local CR frame determination among the three considered approaches. In this case, the proposed force correction approach, or the use of conventional projection matrix, is necessary to ensure the computational accuracy.

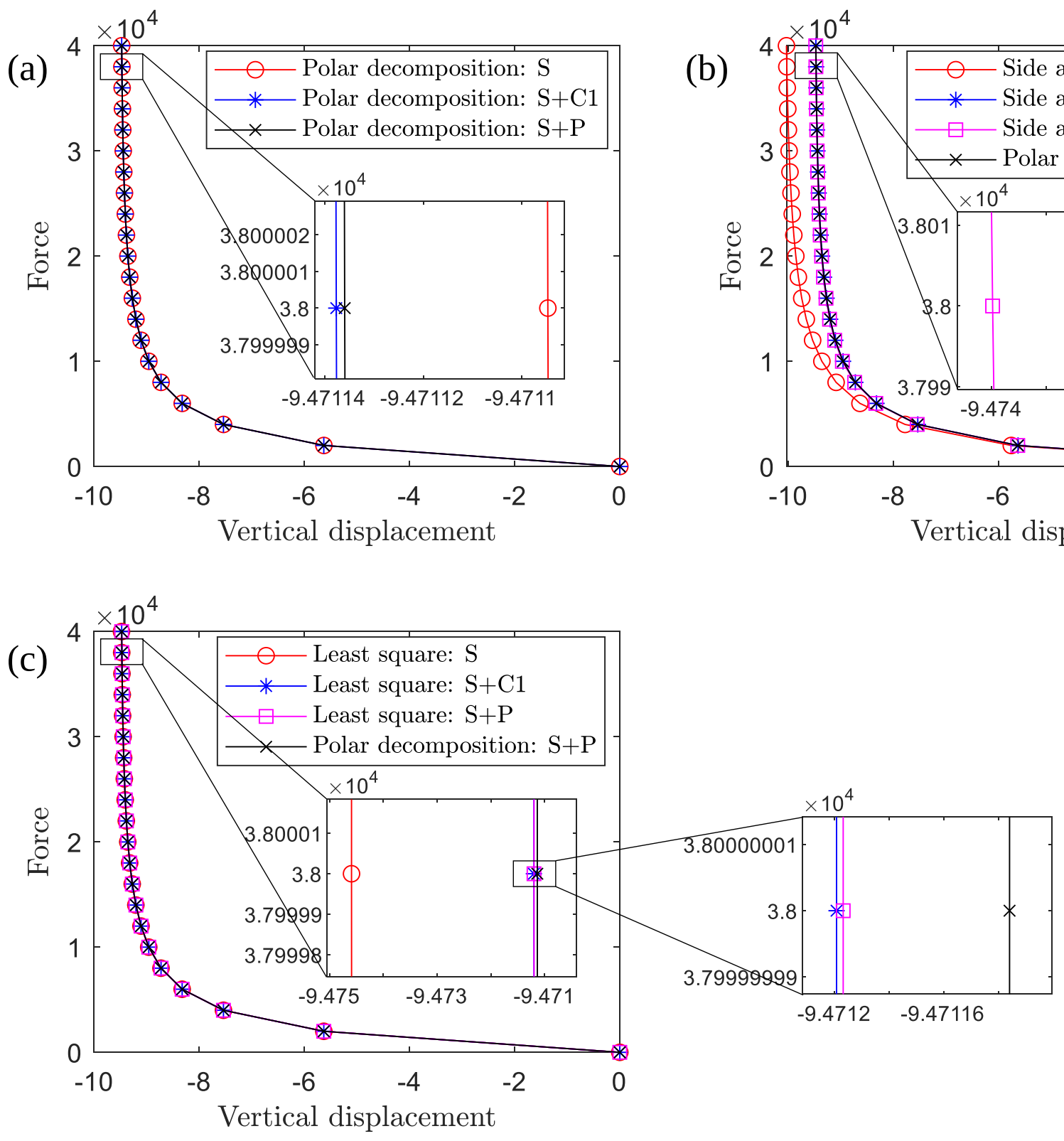

Fig. 11 Evolution curves of the vertical displacement of the loading point with three choices of local CR frame.

### 4.2. Spatial cantilever structure

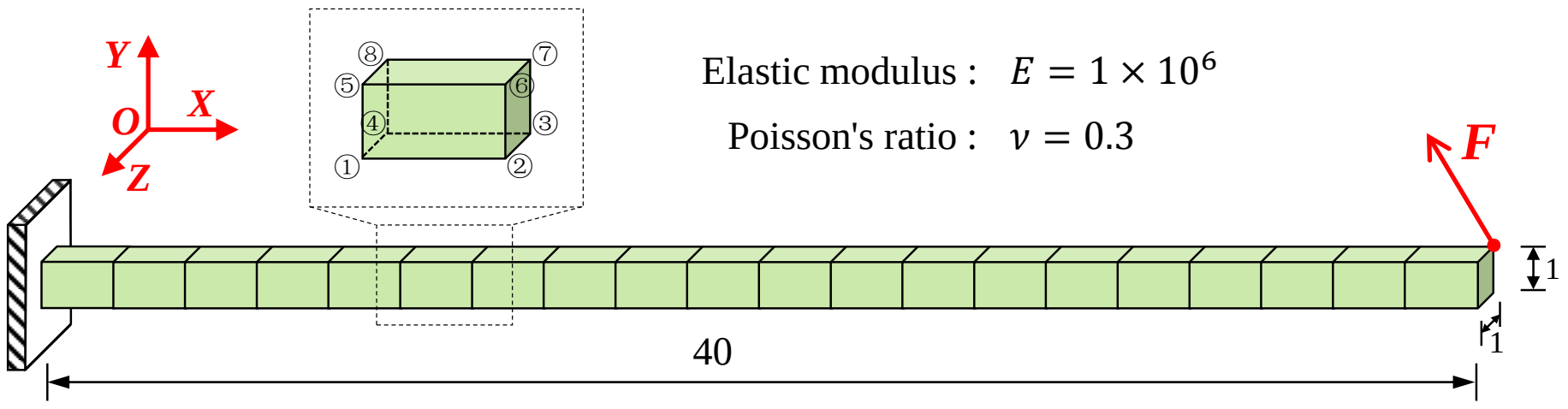

Fig. 12 Spatial cantilever structure: geometrical and material parameters, and FEM model.

We next test the performance of the developed framework for solid elements. As shown in Fig. 12, the example involves a spatial cantilever structure, which is clamped at one end and is subjected to a spatial concentrated force at one corner of the other end. The maximum force vector is set as $\boldsymbol{F}_{\text{max}} = [-1000; 200; 200]$. We model the structure with twenty 8-node spatial solid elements. The results are obtained using 20 equal loading steps.

Fig. 13 presents the curves of the *x*/*y*/*z* displacements of the loading point for two choices of local CR frame with different internal force calculation methods. As can be seen, for both approaches the S+C1 and the S+P methods yield similar results, while the S method yields totally different results, demonstrating that the effect of using the proposed force correction approach of importance. To gain an insight into, Fig. 14 presents the final deformed configurations of the structure for different cases. The result obtained by ANSYS (using solid-45 elements with KEYOPT(1)=1) is also plotted in order to highlight the correctness of the present studies.

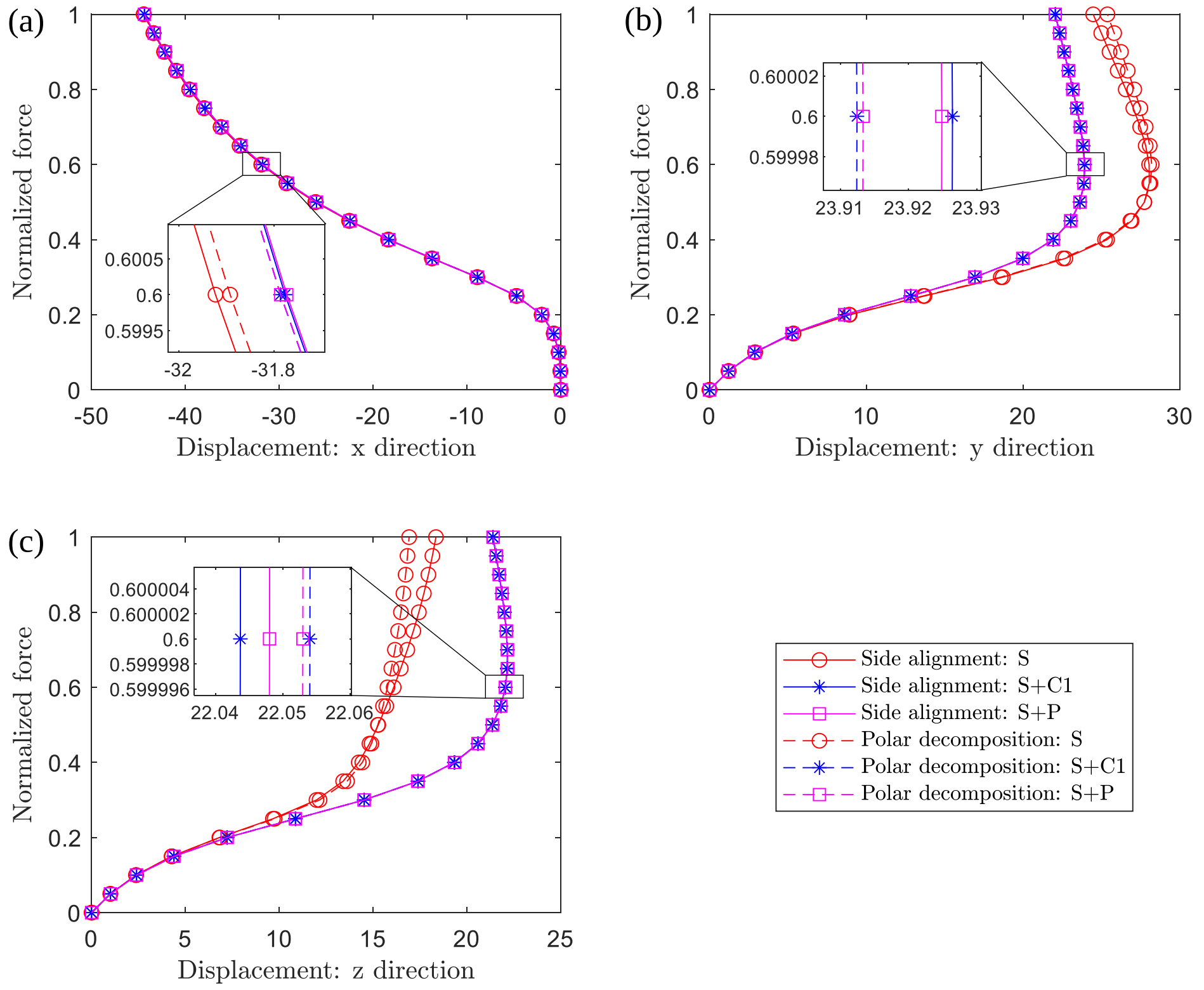

Fig. 13 Evolution curves of the displacements of the loading point for different cases

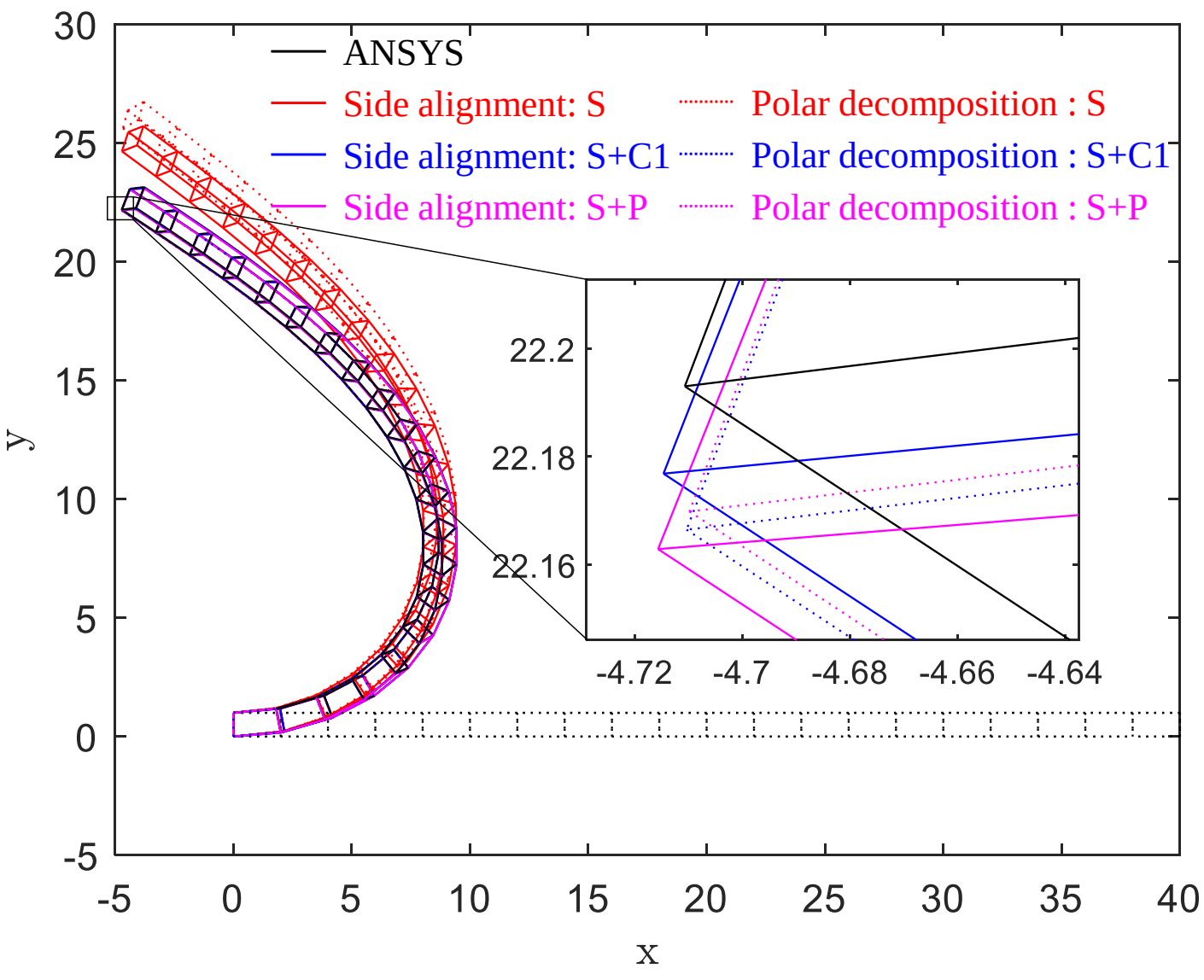

Fig. 14 Deformed configurations at the maximum load for different cases.

### 4.3. Spatial angle frame

We then test the performance of the developed framework for beam elements. The example involves a spatial frame which was first studied in [46]. As shown in Fig. 15. the structure consists of three straight beams connected at right angles. It is clamped at one end and is subjected to a spatial concentrated force at the central point of the other end. The maximum force vector is set as $\boldsymbol{F}_{\max} = -[5;0;5]$, with such a value the displacement is of the same order of magnitude as the length of the frame. We model the structure with twelve 2-node beam elements. The results are obtained using 20 equal loading steps.

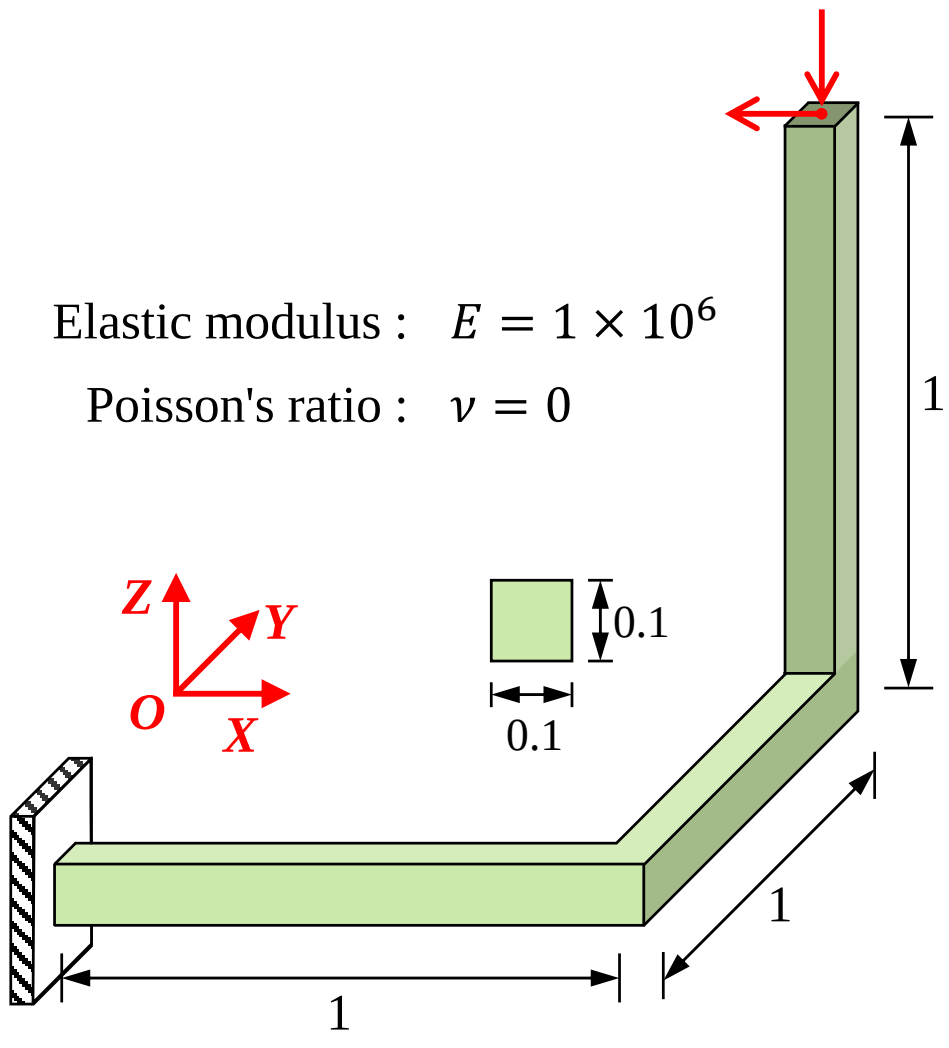

Fig. 15 Spatial angle frame: geometrical and material parameters.

Fig. 16 presents the curves of the *x*/*y*/*z* displacements of the loading point for different internal force calculation methods, under the previously summarized choice of the local CR frame for beam element. It is shown that all internal force calculation methods yield almost identical results. This is due to the fact that the present choice of the local CR frame is sufficiently accurate to reflect the rigid body motion of the beam elements. More calculations show that for the internal force calculated solely with S method, the maximum magnitude of the parasitic unbalance moment of the element internal force is just at the order of $1\times10^{-4}$. In this situation, the effect of using the present force correction approaches or the nonlinear projection matrix is limited. However, we would like to explore the differences of the obtained results from a micro-perspective. From the enlarged view of Fig. 16, one can found that compared with the result obtained solely with S method, those obtained with various force correction approaches are more close to the result obtained with conventional nonlinear projection matrix, in particular for the S+C3 method. Fig. 17 presents the final deformed configurations of the structure. The configuration obtained by ANSYS (using beam-4 elements) is also plotted in order to highlight the correctness of those obtained results.

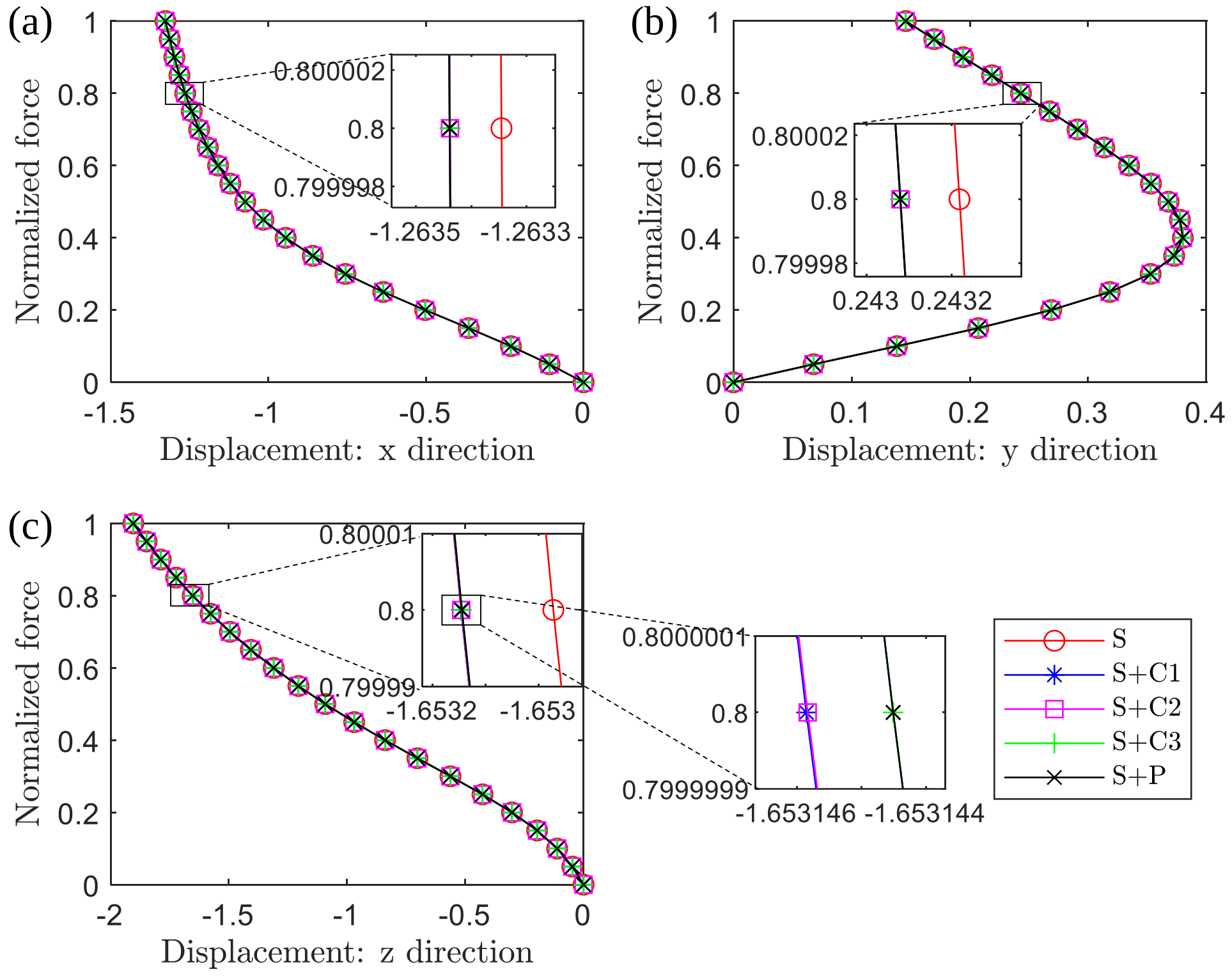

Fig. 16 Evolution curves of the displacements of the loading point for different cases.

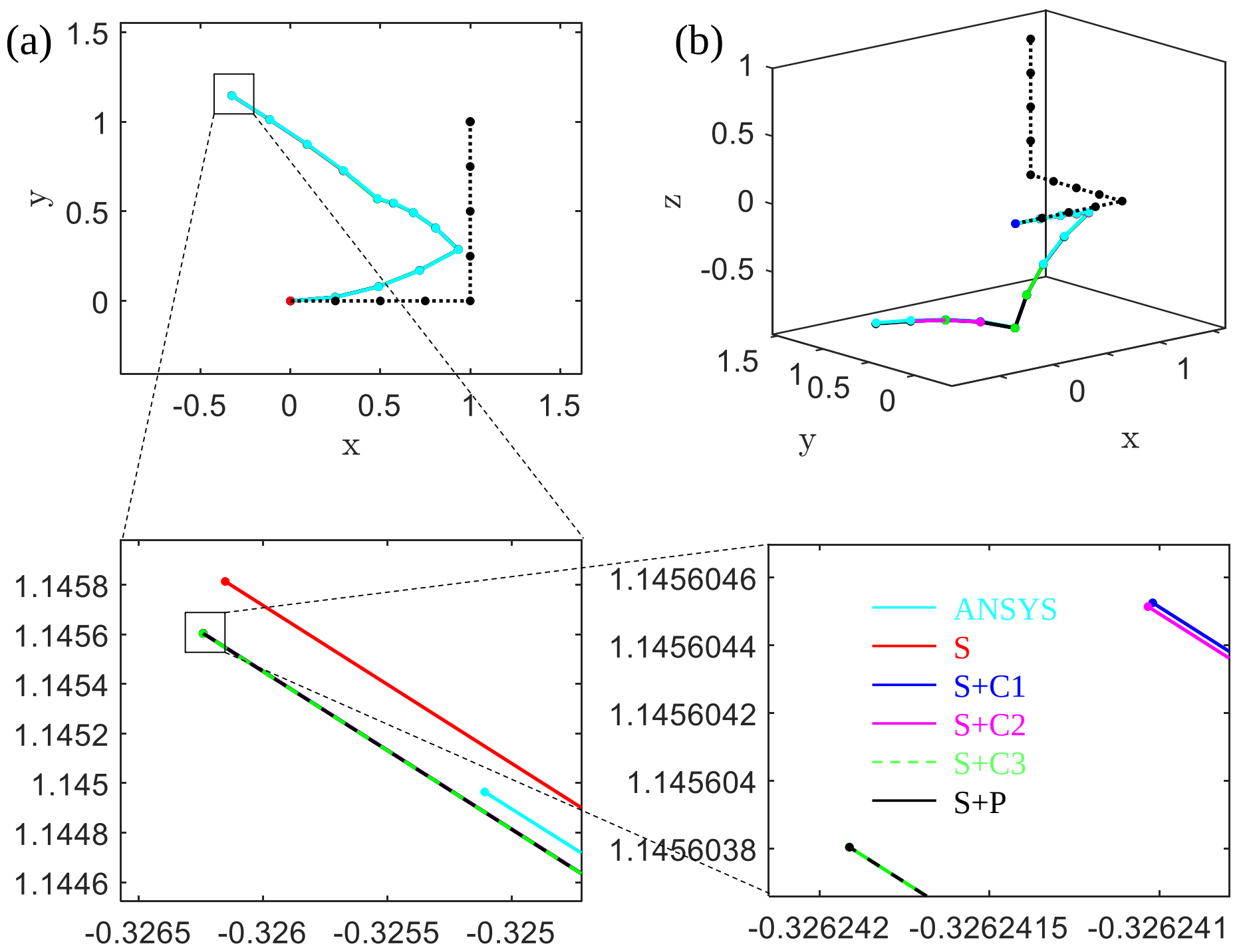

Fig. 17 Deformed configurations at the maximum load for different cases: (a) a top view; (b) a perspective view.

## 4.4. L-shaped frame

In the following examples, we focus on testing of the performance of the developed framework for shell elements, in particular for the quadrilateral shell element, using some benchmark examples. The first one concerns with lateral buckling analysis of a L-shaped frame. This classical problem has been investigated in many existing studies [7, 34, 47]. As shown in Fig. 18, the L-shaped frame is clamped at the bottom end, and is subjected to an in-plane load at the free end. The frame is driven to the buckling mode by a small perturbation load ($1\times10^{-3}$) initially applied at the free end normal to the plane of frame. It should be noted that the cross section possesses an extreme slenderness, i.e., thickness/height =1/50, which leads to a buckling behavior of the deformation. We consider the applied force varying from 0 to 2, with equal load increment of magnitude $5\times10^{-3}$.

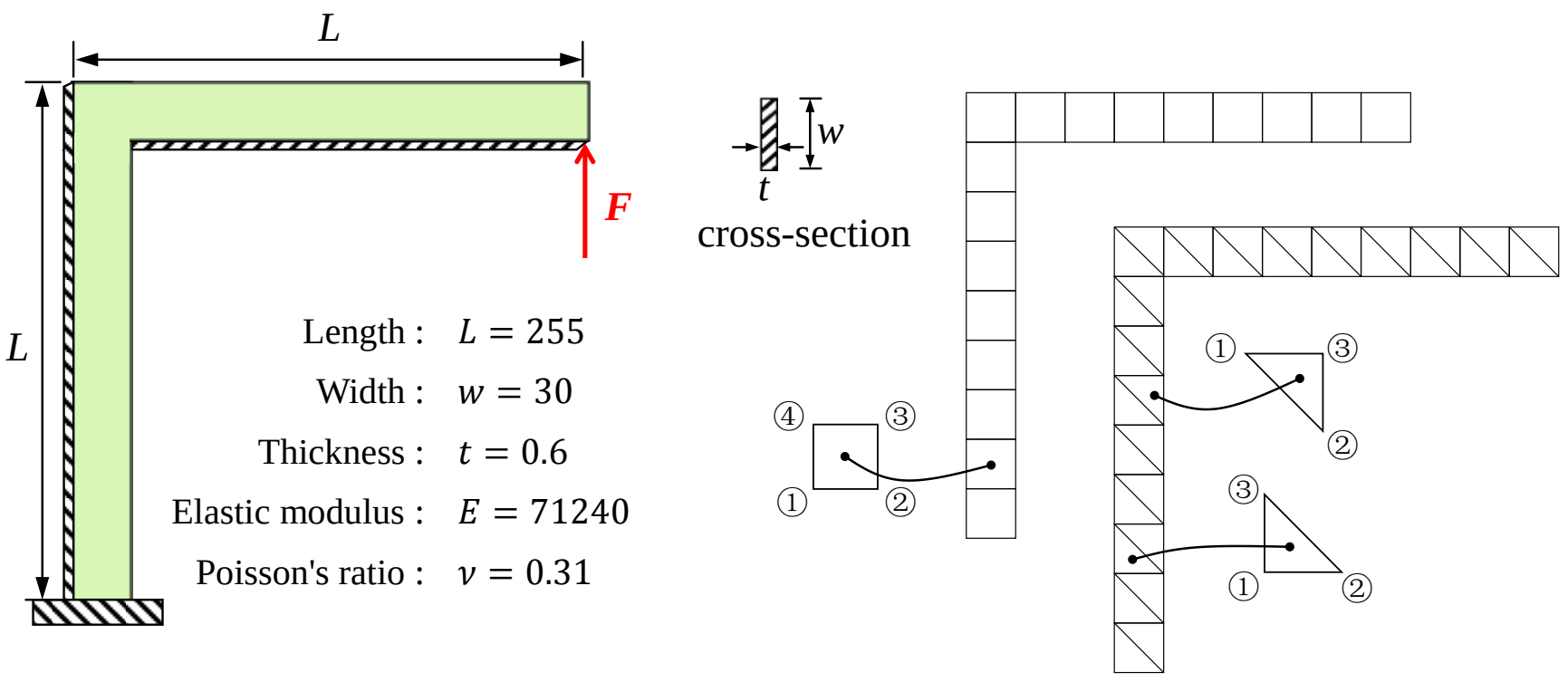

Fig. 18 L-shaped frame: geometrical and material parameters, and FEM models.

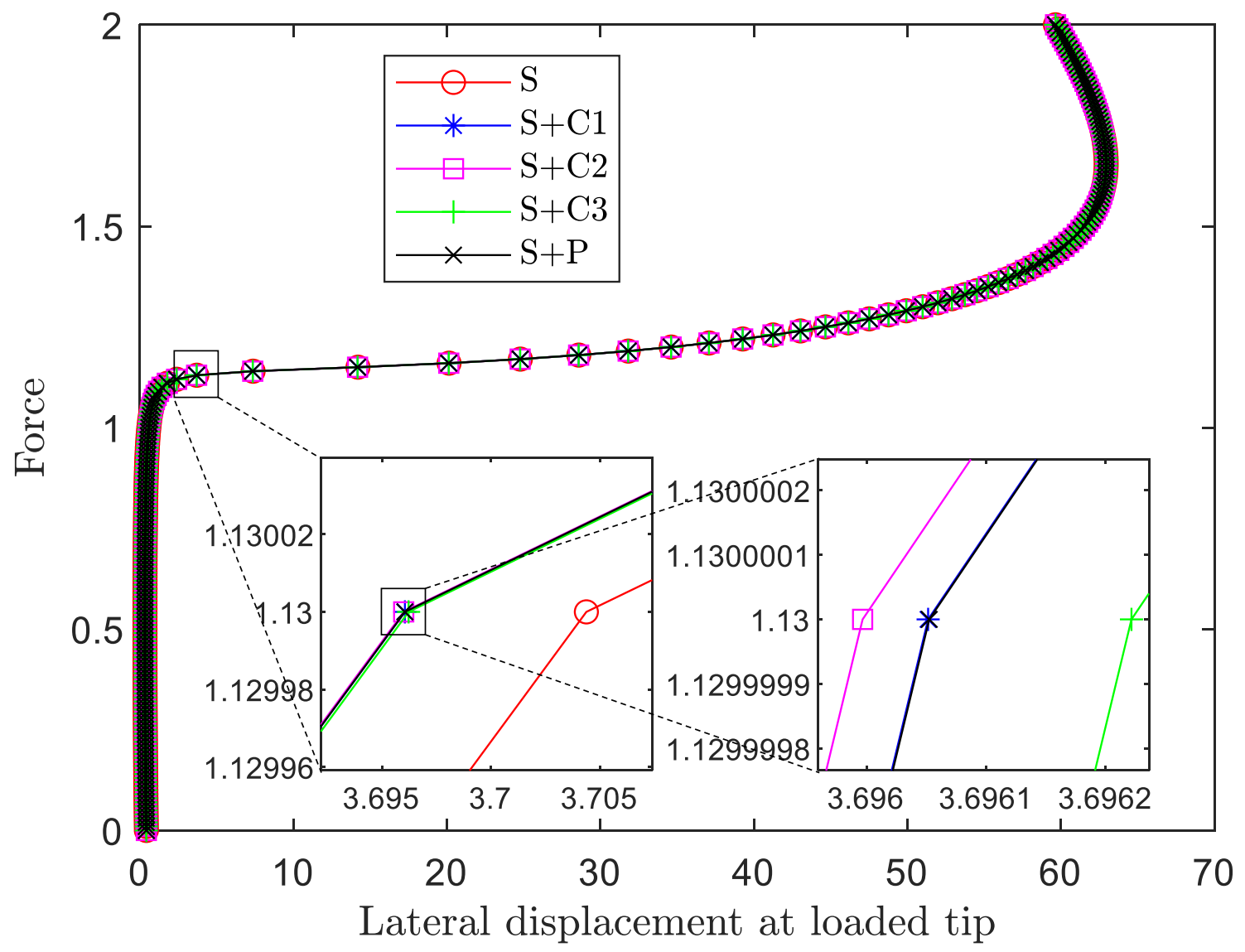

Fig. 19 Evolution curves of the lateral displacement of the loading point for different internal force calculation methods.

We first conduct the analysis with triangle shell elements. Fig. 19 presents the evolution of the lateral displacement at the loading point for different internal force calculation methods. It can be found that all internal force calculation methods yield almost identical results. This is due to the fact that for this example the elements are subjected to small membrane strains, and also that no warping effects for the triangle shell element, the current side alignment approach is sufficiently accurate to capture the rigid rotation of the elements. In this situation, the effect of using the force correction approaches or using the nonlinear projection matrix is limited. Nevertheless, from the enlarged view of Fig. 19, one can found that compared with the result obtained solely with S method, those obtain with the proposed force correction approaches are more close to the result obtained with conventional nonlinear projection matrix. We have also conduct the analysis with ANSYS (using shell 63 element with KEYOPT (3)=1). The outcome (not presented to save space) is fairly close with the above obtained results, demonstrating the correctness of the present results.

Apart from this particular example, we have also tested many other classical benchmark problems with *triangle* shell elements. A general conclusion is that for many examples documented in existing literature the triangle elements typically possess small membrane strains, and the internal force calculated solely by the S method is able to yield a reasonably accurate result comparing with those by the S+P method. This implies that in many cases the effect of using projection matrix for triangle shell element is exaggerated for *triangle* shell elements. In the following examples, we therefore limit our comparisons for only quadrilateral shell element.

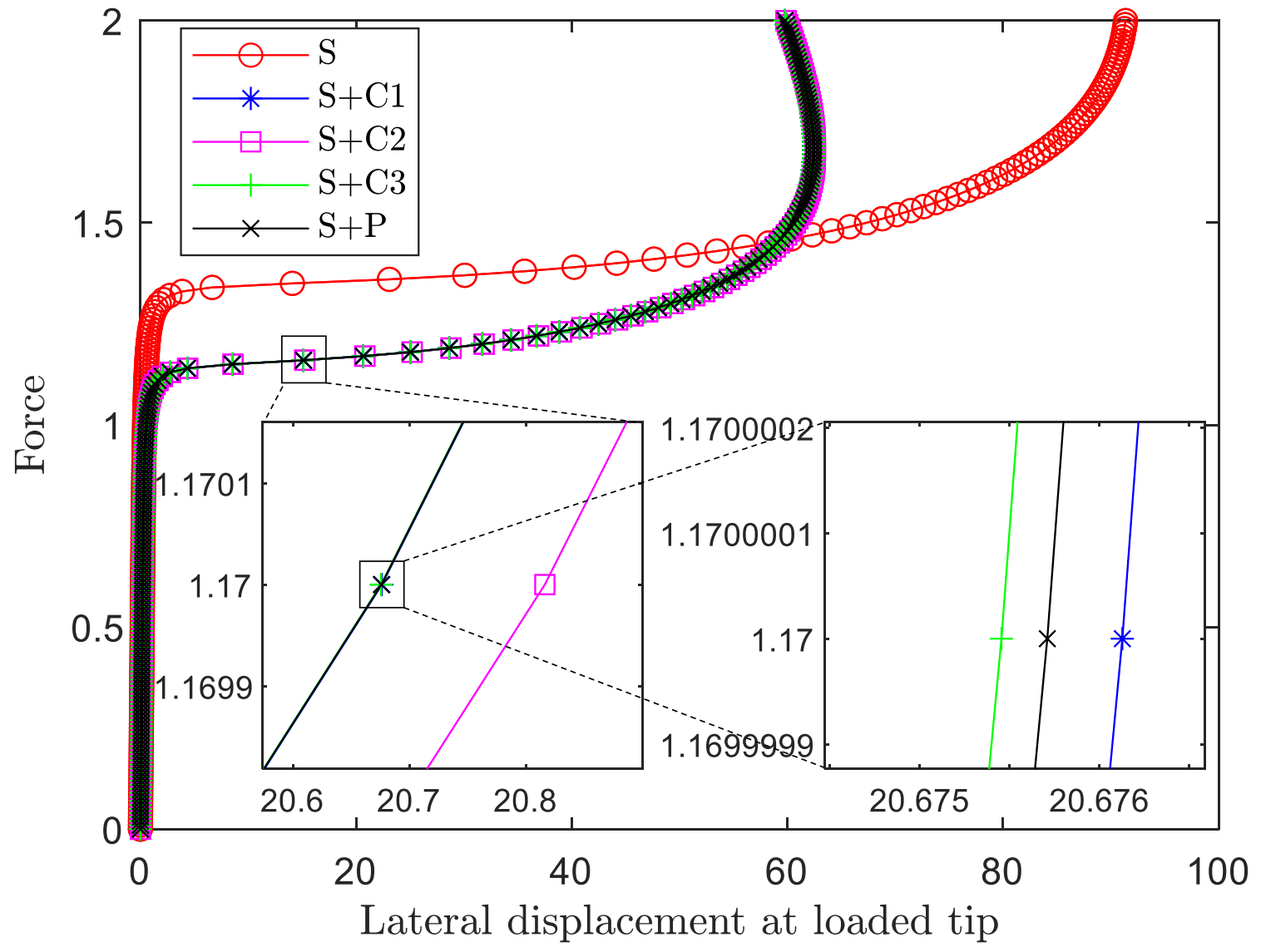

Fig. 20 Evolution curves of the lateral displacement of the loading point for different internal force calculation methods.

Fig. 20 presents the evolution of the lateral displacement at the loaded tip for the analysis with quadrilateral shell elements. It can be found that the results obtained with the proposed three force correction approaches agree well with those obtained with conventional nonlinear projection matrix, while the internal force calculated solely with S method yields a completely different result. These outcomes demonstrate that the effect of using the proposed force correction approaches or using the nonlinear projection matrix in this example is of great importance. Fig. 21 presents the final deformed configurations of the structure for different force calculation methods. The configuration obtained by ANSYS (using shell 63 element with KEYOPT (3)=1) is also plotted in order to highlight the correctness of the present result.

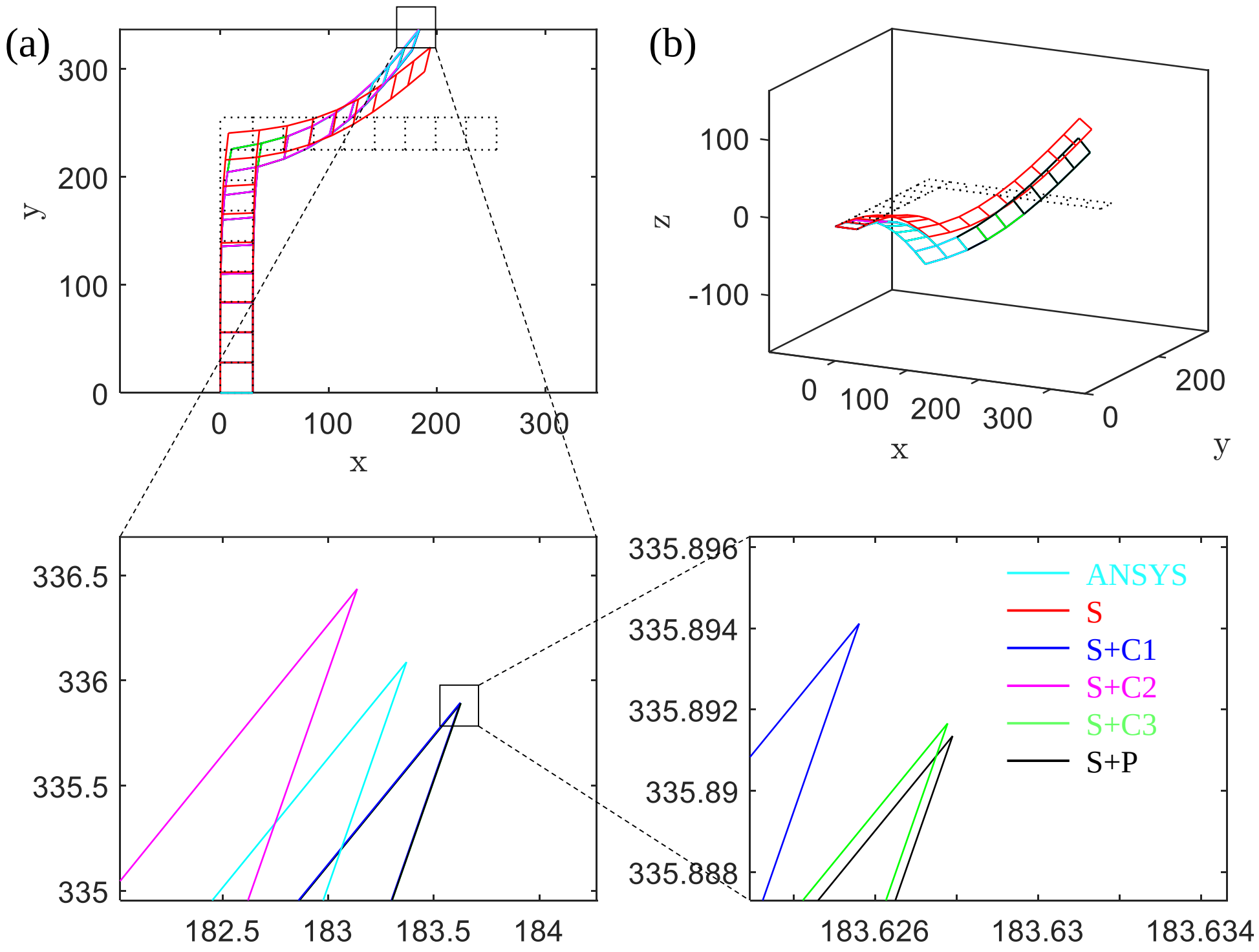

Fig. 21 Deformed configurations at the maximum load for different cases: (a) a top view; (b) a perspective view.

## 4.5 Pre-twisted beam

This example shown in Fig. 22 involves a pre-twisted beam, which was first investigated by MacNeal and Harder [48]. Here, we use a version of the problem adopted in [12] where the thickness of the beam is set as 0.05. We conduct the analysis using quadrilateral shell elements with $4\times 24$ mesh. The elements are initially warped. The maximum force magnitude is set as $\boldsymbol{F}_{\max} = 60$, and 20 equal loading steps are used.

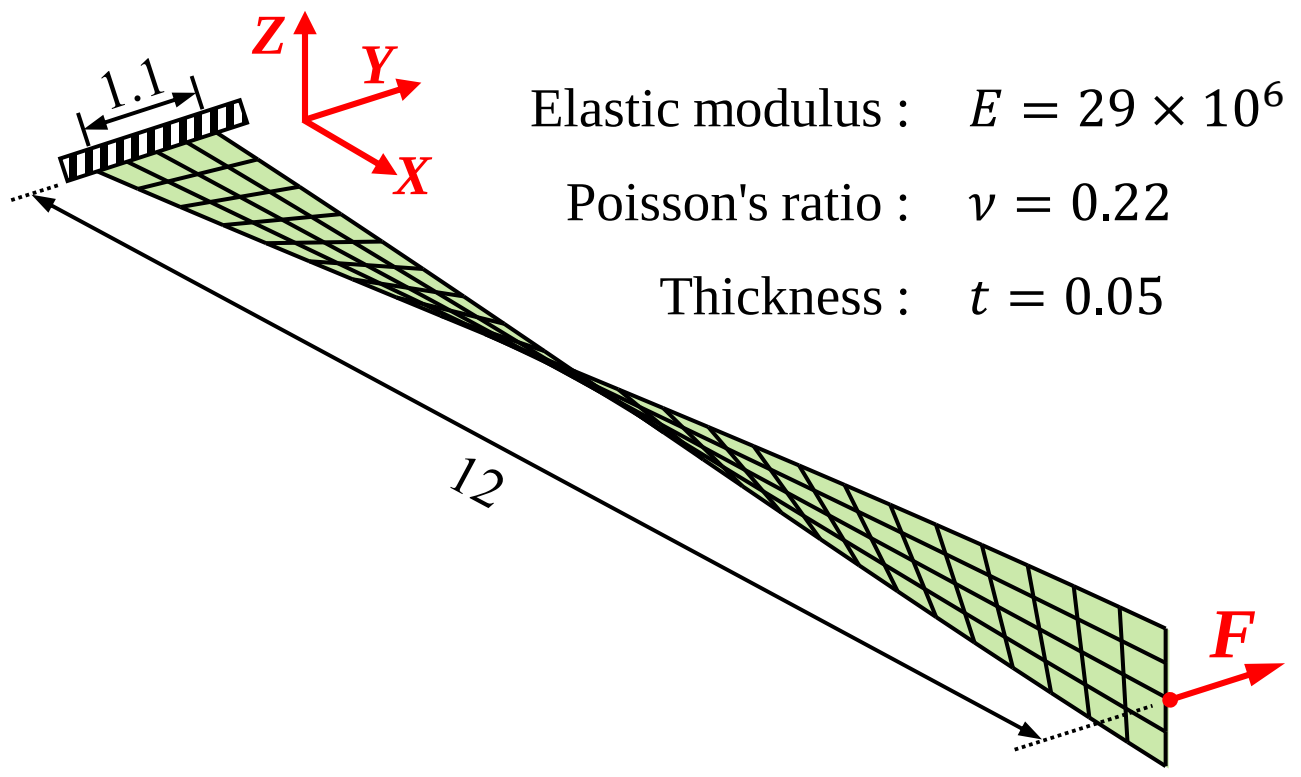

Fig. 22 Pre-twisted beam: geometrical and material parameters, and FEM model.

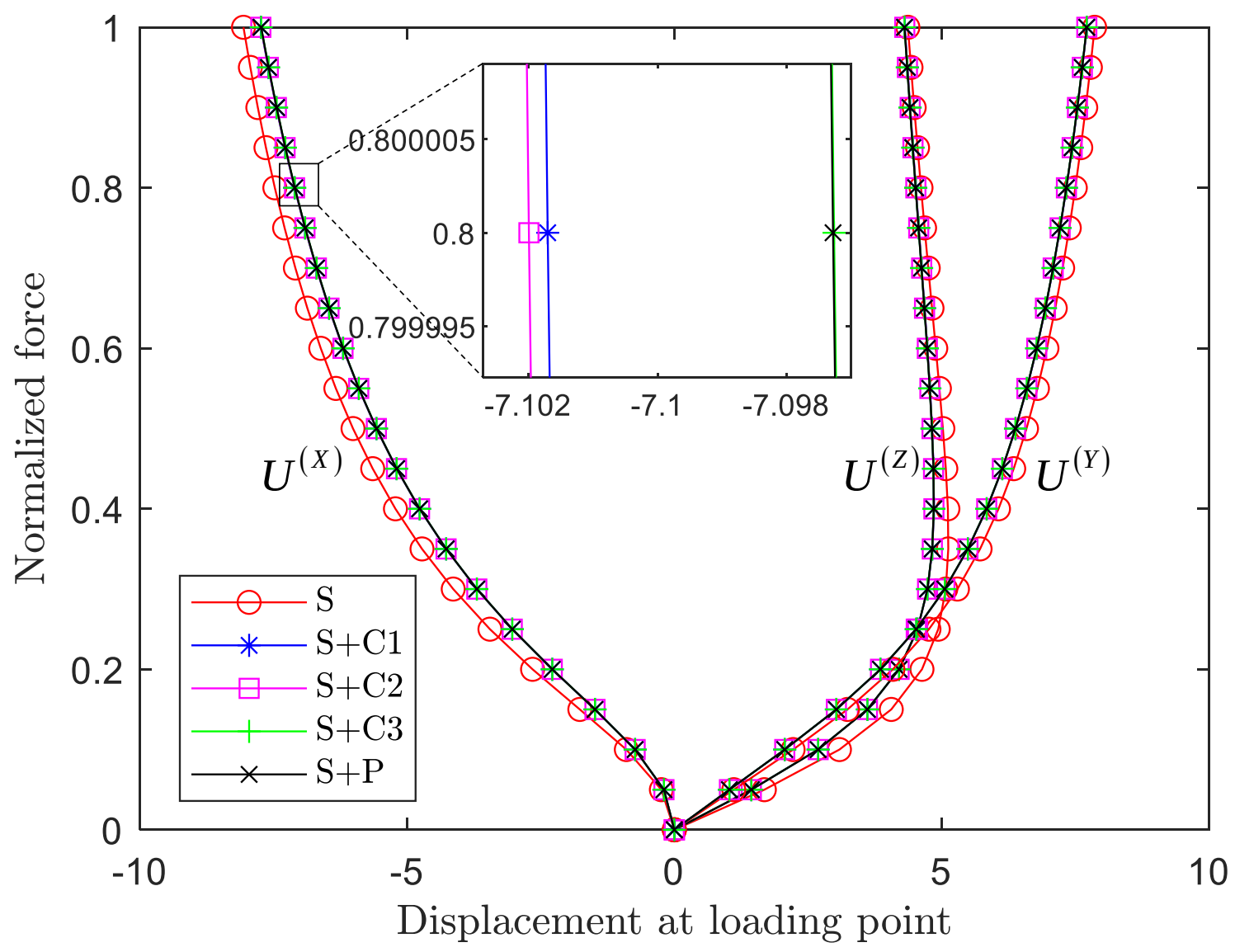

Fig. 23 Evolution curves of the displacement of the loading point for different internal force calculation methods.

The displacements along the *x*/*y*/*z* directions of the loading point are plotted in Fig. 23. The results obtained with the proposed three force correction approaches agrees well with those obtained with conventional nonlinear projection matrix. For this example, it appears that the results obtained by S+C3 method is even closer with those obtained by S+P method. Fig. 24 presents the final structural deformed configurations for different force calculation methods. The configuration obtained by ANSYS (using shell 181 element with KEYOPT (3)=2) is also presented.

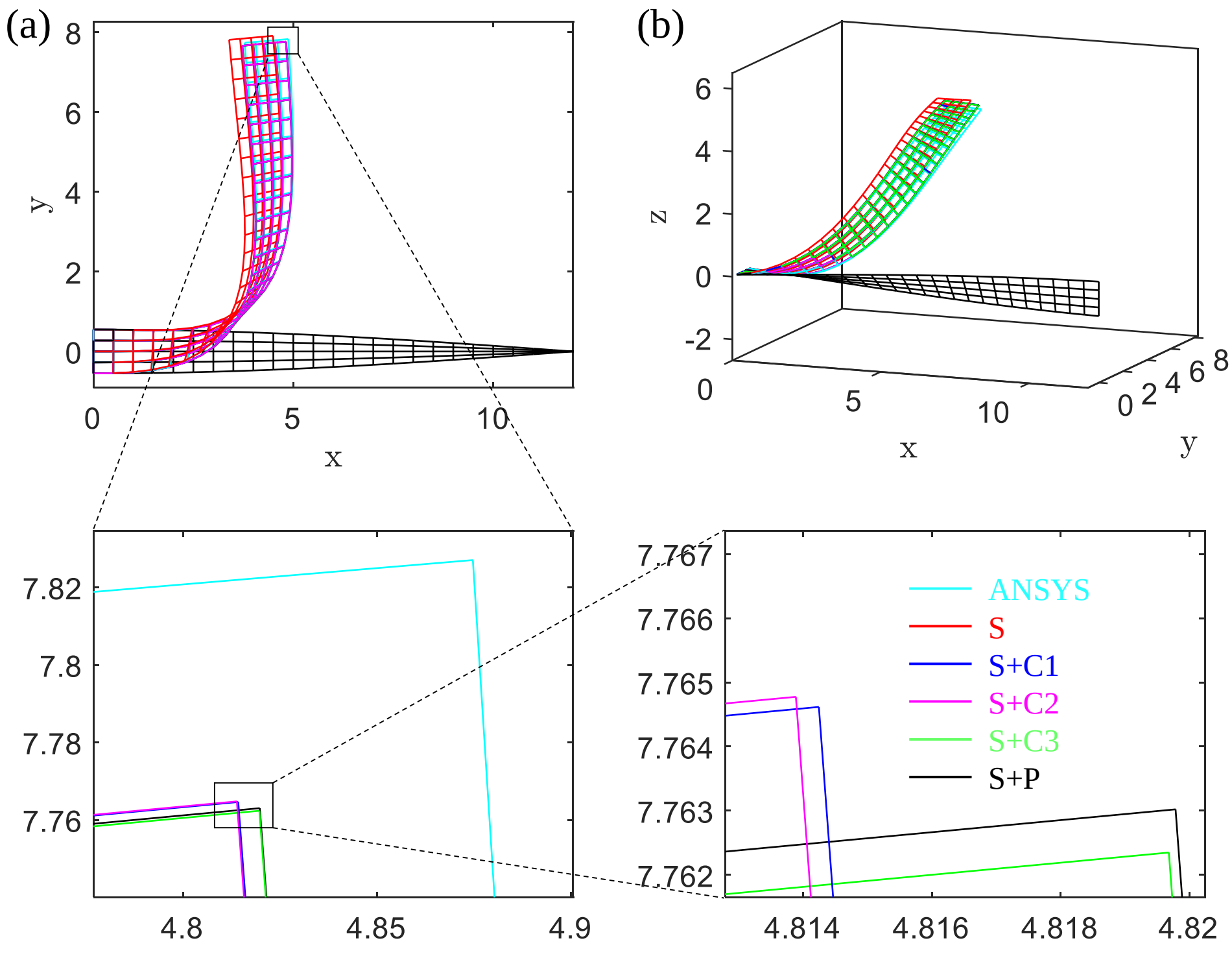

Fig. 24 Deformed configurations at the maximum load for different internal force calculation methods: (a) a top view; (b) a perspective view.

4.6 Slit annular plate

In this example we consider a slit annular plate, which has been investigated in a large number of existing works [49, 50]. As shown in shown in Fig. 25, a line force is applied at one end of the slit while the other end of the slit is fully clamped. The maximum line force magnitude (force/length) is taken as 0.8. We conduct the analysis using quadrilateral shell element with $6 \times 30$ mesh. The results are obtained with 50 equal loading steps.

Elastic modulus : $E = 21 \times 10^6$ Outer/Inner radius : $R_o = 10; R_i = 6$

Poisson's ratio : $\nu = 0$ Thickness : $t = 0.03$

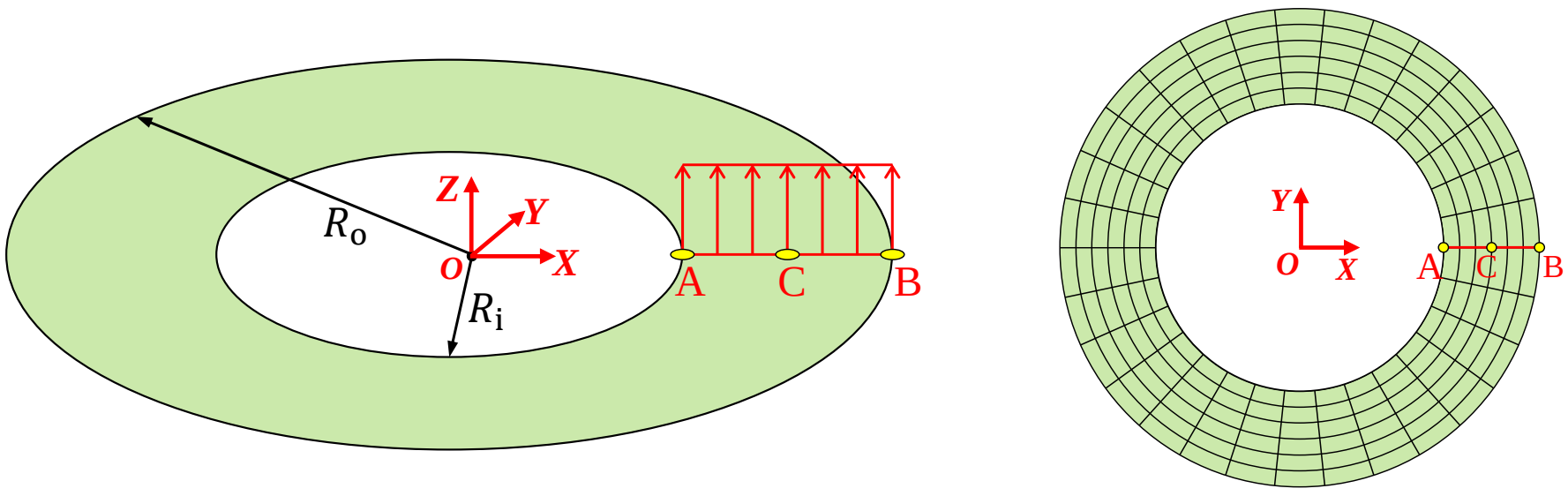

Fig. 25 Slit annular plate: geometrical and material parameters, and FEM model.

Fig. 26 plots the evolution of the displacements of the central point (C) of the slit. Again, the results obtained with the proposed three force correction approaches agree well with those obtained with conventional nonlinear projection matrix, while the internal force calculated solely with S method yields completely different results, demonstrating that the effect of using the proposed force correction approaches or using the nonlinear projection matrix is of significant importance in this analysis. Fig. 27 gives the structural deformed configurations under the maximum load. The configuration obtained by ANSYS (using shell 63 element with KEYOPT (3)=1) is also plotted in order to highlight the correctness of the present results.

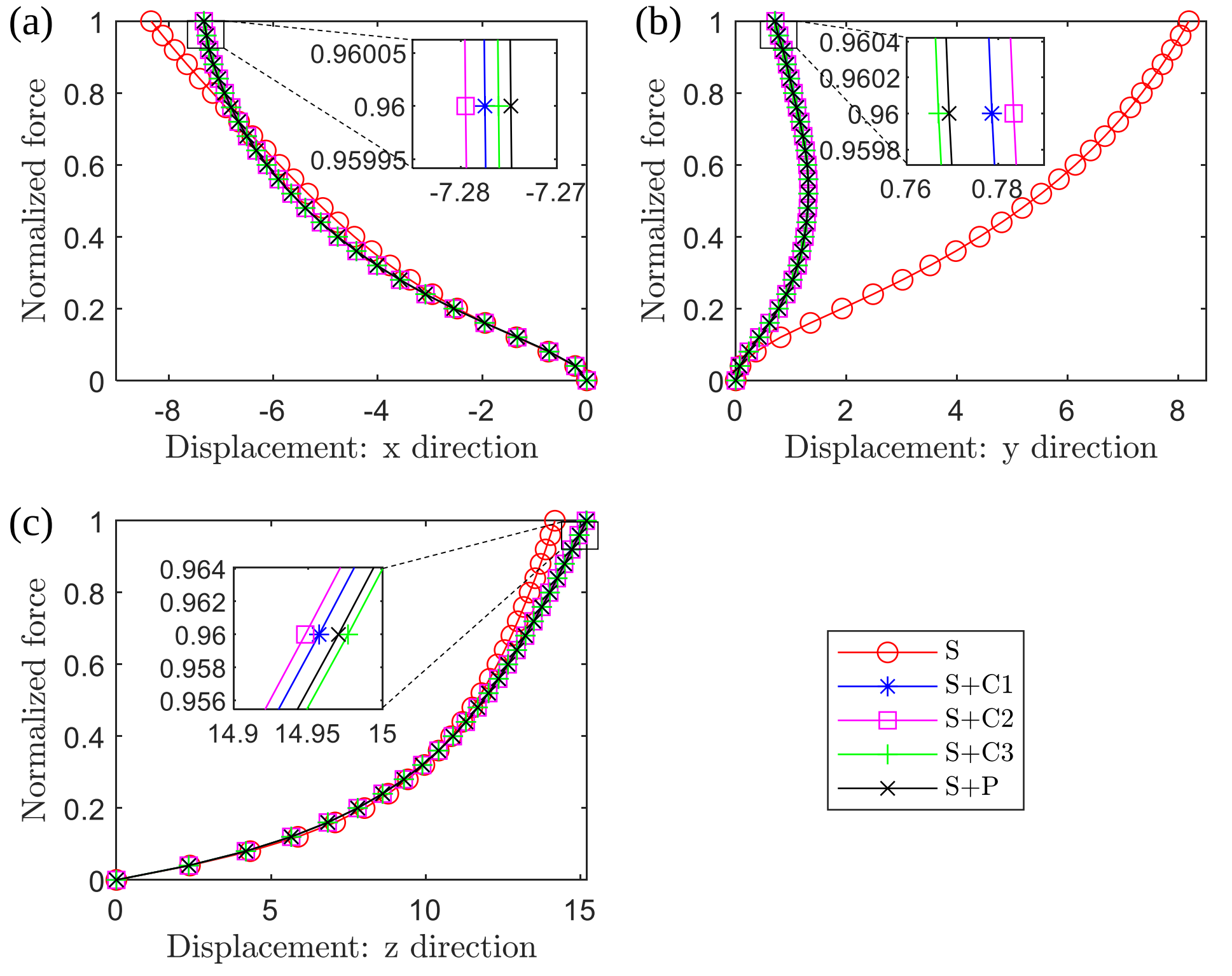

Fig. 26 Evolution curves of the displacements of the central point (C) of the slit for different internal force calculation methods.

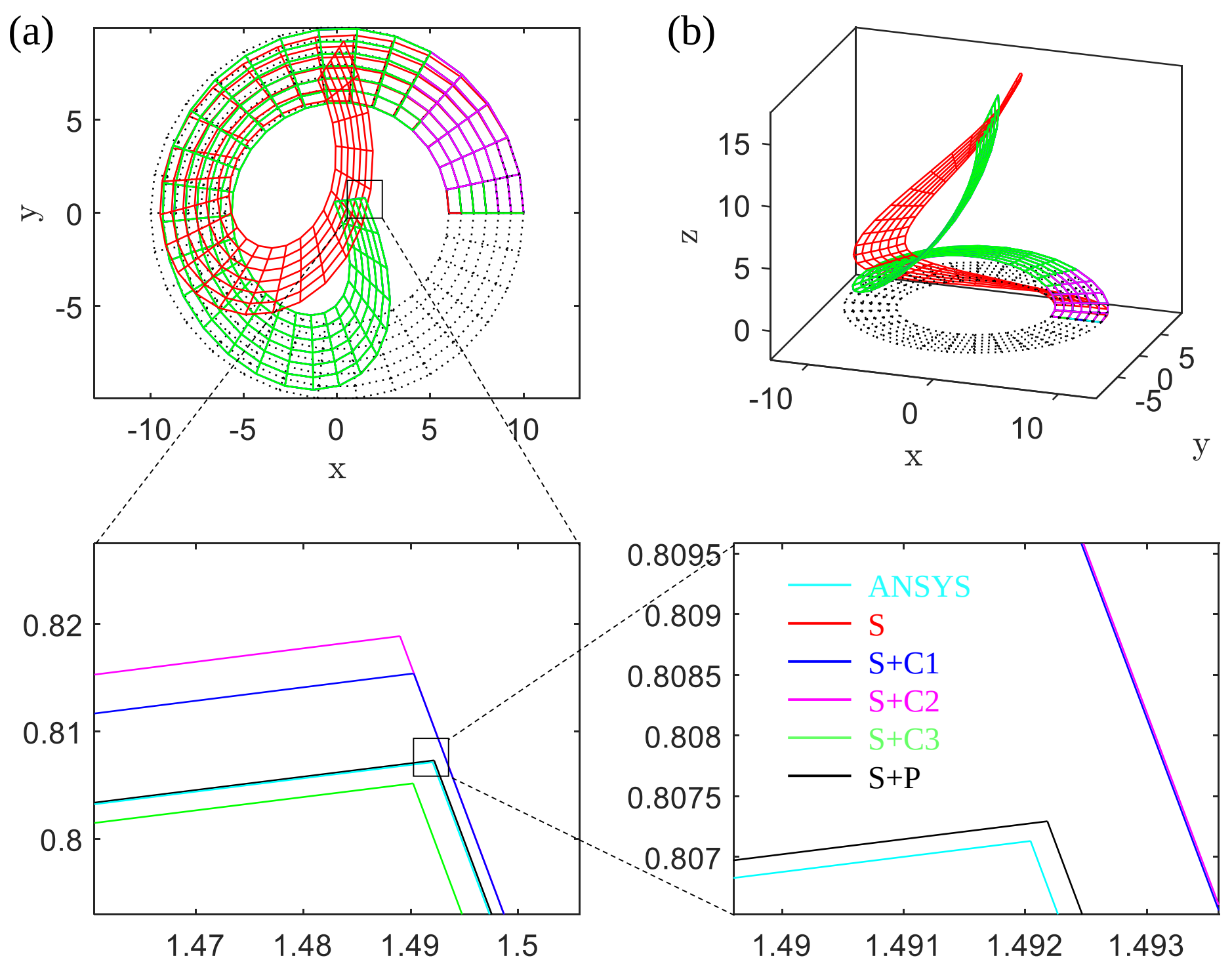

Fig. 27 Deformed configurations at the maximum load for different internal force calculation methods: (a) a top view; (b) a perspective view.

## 5. Conclusions

The use of nonlinear projection matrix in CR analysis was pioneered by Rankin and Nour-Omid in 1990s [6, 7], and has almost became a standard manner for CR formulations deduction over the past thirty years. This matrix however relies heavily on a hysterical and sophisticated derivation of the variations of the local displacements to the global ones, leading to complicated expressions for the internal force vector and the tangent stiffness matrix, which may devalue the simplicity and convenience for the original intention of using CR approach.

This paper begins by making a discussion on existing element independent CR formulation, and the objective is to develop a new and simple framework for general CR analysis that avoids using conventional nonlinear projection matrix. The methodology consists of two steps in the element calculation. The first one is to obtain a preliminary result of the internal force. This is done by following the conventional element-independent CR formulation but drops the terms involving projection matrix and therefore yields simple formulations of the internal force and the tangent stiffness matrix. The second one is a correction step to obtain a new internal force vector that satisfies the element self-equilibrium condition based on the minimum norm correction approach. This step inherits the spirt of using projection matrix but is conducted directly in the global frame, thus avoiding complicated entanglement of local-global rotation and is independent of the choice of the local CR frame used in the analysis. This further leads to a simple and unified formulation for different kinds of elements that can be cooperated in CR framework. Closed formulation of the correction force term as well as the related consistent tangent stiffness matrix are derived. For structural elements, by using different weight matrices, three optional correction approaches are provided. It is also shown that the existing projector matrix under infinitesimal rotation is a special case of the correction approach.

Multiple numerical examples involving various kinds of elements and different choices of element local CR frame are presented to reveal the performance of the proposed framework. The outcomes suggest that in some special examples with some particular choices of local CR frame, using solely S method may be sufficient to obtain an accurate result. In these cases, the effects of using nonlinear projection matrix, as well as using the present force correction approach, is unnecessary. However, for general cases the effects using the present force correction approach are remarkable and the accuracy of the results is comparable with those obtained with conventional nonlinear projection matrix.

The present studies demonstrate that it is possible to analyze structures undergoing large rotations within a general and simpler framework. Although limited to static and linear elastic analyses, it is straightforward that the methodology of force correction can be extended to implicit or explicit dynamic analyses. It is also promising to extend the approach into analyses with material nonlinearity. Testing the performance of the proposed methodology in these applications is a topic of our future work.

**Acknowledgments**

The authors are grateful for the financial support of the National Key Research and Development Plan (2019YFB1706502); the National Natural Science Foundation of China (12002072,11922203); the fellowship of China Postdoctoral Science Foundation (2020M680943).

## Appendix

### A. Formulations of finite rotations

In this Appendix we briefly summarize some basic expressions of finite rotations involved in this work. Interested readers are referred to [51] and the Appendix of [38] for more details.

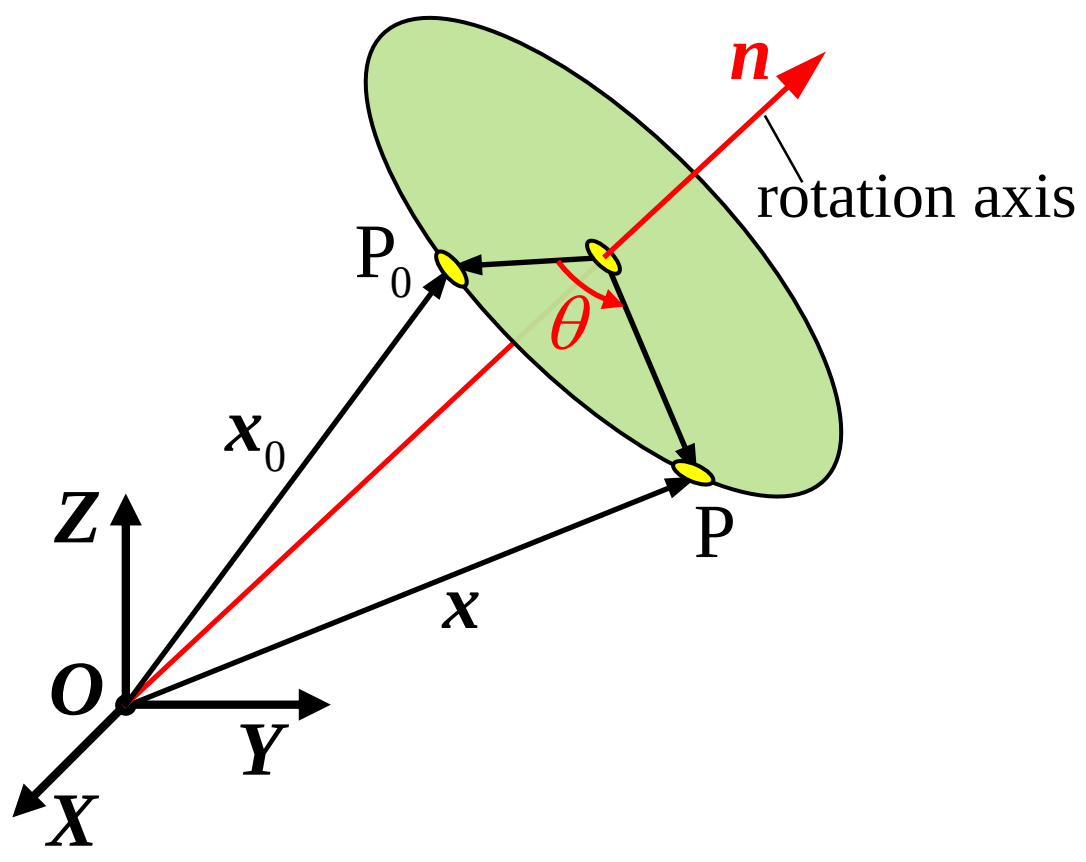

Fig. 28 Spatial finite rotation and rotation vector.

As shown in Fig. 28, in three-dimensional position vector $\boldsymbol{x}_0$ can be rotated into a new position vector $\boldsymbol{x}$ by a rotation matrix $\boldsymbol{R}$, which implies

$$\boldsymbol{x} = \boldsymbol{R}\boldsymbol{x}_0 \tag{54}$$

The rotation matrix $\boldsymbol{R}$ is a 3×3 orthonormal matrix, i.e., $\boldsymbol{R}^{-1} = \boldsymbol{R}^{\mathrm{T}}$ and det($\boldsymbol{R}$)=1. The rotation matrix $\boldsymbol{R}$ can be parameterized using the rotation vector that is defined by

$$\boldsymbol{\theta} = \theta \boldsymbol{n} \tag{55}$$

where $\boldsymbol{n}$ represents the unit vector defining the rotation axis and $\theta$ represents the rotation angle. Using the rotation vector $\boldsymbol{\theta}$, the rotation matrix can be expressed as

$$\boldsymbol{R} = \exp(\boldsymbol{\Theta}) = \boldsymbol{I}_3 + \frac{\sin\theta}{\theta}\boldsymbol{\Theta} + \frac{1-\cos\theta}{\theta^2}\boldsymbol{\Theta}^2 \tag{56}$$

where $\boldsymbol{\Theta}$ represents the spin of the rotation vector which means $\boldsymbol{\Theta} = \mathrm{spin}(\boldsymbol{\theta})$ or inversely $\boldsymbol{\theta} = \mathrm{axial}(\boldsymbol{\Theta})$.

For structural elements in CR analyses, the extraction of the rotation angle $\theta$ and the unit rotation axis $\boldsymbol{n}$ is often required for a given rotation matrix $\boldsymbol{R}$, and can be obtained from

$$\cos\theta = \frac{1}{2}\left(\mathrm{trace}(\boldsymbol{R}) - 1\right); \boldsymbol{n} = \frac{\mathrm{axial}\left(\boldsymbol{R} - \boldsymbol{R}^{\mathrm{T}}\right)}{2\sin\theta} \tag{57}$$

The above formulas are prone to numerical instability for angles near 0, $\pm\pi$, etc., because $\sin\theta$ vanishes. A robust algorithm is given by Spurrier [52] in the language of quaternions.

The theoretical formula for the logarithm of a rotation matrix can also be obtained from

$$\boldsymbol{\Theta} = \log_{\mathrm{e}}(\boldsymbol{R}) = \frac{\arcsin(\tau)}{2\tau}\left(\boldsymbol{R} - \boldsymbol{R}^{\mathrm{T}}\right); \tau = \frac{1}{2}\left\|\mathrm{axial}\left(\boldsymbol{R} - \boldsymbol{R}^{\mathrm{T}}\right)\right\| \tag{58}$$

It is worth mentioning that the formula fails, however, outside the range $[-\pi/2 \le \theta \le \pi/2]$ and is numerically unstable near $\theta = 0$.

**B. Derivation of Eq. (6)**

In this Appendix, we derive Eq. (6) following the methodology of Rankin and Nour-Omid [6, 7].

First, the term involving the variation of the orthogonal matrix can be written at the node level, as

$$\frac{\partial\left(\mathrm{diag}\left(\boldsymbol{R}^{\mathrm{T}}\right)\boldsymbol{x}\right)}{\partial\boldsymbol{u}}=\begin{bmatrix}\dfrac{\partial\left(\boldsymbol{R}^{\mathrm{T}}\boldsymbol{x}_1\right)}{\partial\boldsymbol{u}}\\ \vdots\\ \dfrac{\partial\left(\boldsymbol{R}^{\mathrm{T}}\boldsymbol{x}_N\right)}{\partial\boldsymbol{u}}\end{bmatrix} \tag{59}$$

Each component in (59) can be derived using the differential relationship

$$\mathrm{d}\boldsymbol{R}^{\mathrm{T}}\boldsymbol{x}_i=\frac{\partial\left(\boldsymbol{R}^{\mathrm{T}}\boldsymbol{x}_1\right)}{\partial\boldsymbol{u}}\mathrm{d}\boldsymbol{u} \tag{60}$$

By noting that the orthogonal matrix $\boldsymbol{R}$ can be expressed using a skew-symmetric $\boldsymbol{\Omega}$ or its axial vector $\boldsymbol{\omega}$, i.e., $\boldsymbol{R}=\exp\left(\boldsymbol{\Omega}\right)=\exp\left(\mathrm{spin}\left(\boldsymbol{\omega}\right)\right)$, the variation of the orthogonal matrix thus yields

$$\mathrm{d}\boldsymbol{R}=\mathrm{spin}\left(\mathrm{d}\boldsymbol{\omega}\right)\boldsymbol{R} \tag{61}$$

where $\mathrm{spin}\left(\mathrm{d}\boldsymbol{\omega}\right)$ means the infinitesimal rotations of the base vectors of the orthogonal matrix. Noting the relationships $\mathrm{spin}\left(\mathrm{d}\boldsymbol{\omega}\right)^{\mathrm{T}}=-\mathrm{spin}\left(\mathrm{d}\boldsymbol{\omega}\right)$ and $\mathrm{spin}\left(\mathrm{d}\bar{\boldsymbol{\omega}}\right)=\boldsymbol{R}^{\mathrm{T}}\mathrm{spin}\left(\mathrm{d}\boldsymbol{\omega}\right)\boldsymbol{R}$ hold, for a particular node, we have

$$\begin{aligned}\mathrm{d}\boldsymbol{R}^{\mathrm{T}}\boldsymbol{x}_i&=-\boldsymbol{R}^{\mathrm{T}}\mathrm{spin}\left(\mathrm{d}\boldsymbol{\omega}\right)\boldsymbol{x}_i=-\mathrm{spin}\left(\mathrm{d}\bar{\boldsymbol{\omega}}\right)\bar{\boldsymbol{x}}_i\\&=\mathrm{spin}\left(\bar{\boldsymbol{x}}_i\right)\mathrm{d}\bar{\boldsymbol{\omega}}=\mathrm{spin}\left(\bar{\boldsymbol{x}}_i\right)\sum_{j=1}^{N}\frac{\partial\bar{\boldsymbol{\omega}}}{\partial\bar{\boldsymbol{u}}_j}\frac{\partial\bar{\boldsymbol{u}}_j}{\partial\boldsymbol{u}_j}\mathrm{d}\boldsymbol{u}_j\\&=-\mathrm{spin}\left(\bar{\boldsymbol{x}}_i\right)^{\mathrm{T}}\sum_{j=1}^{N}\frac{\partial\bar{\boldsymbol{\omega}}}{\partial\bar{\boldsymbol{u}}_j}\boldsymbol{R}^{\mathrm{T}}\mathrm{d}\boldsymbol{u}_j\end{aligned} \tag{62}$$

With Eqs. (59),(60) and (62), one can verify that Eq. (6) is satisfied.

**C. Complex step finite difference**

The conventional nonlinear projection matrix typically takes the form $\bar{\boldsymbol{P}}=\boldsymbol{I}-\bar{\boldsymbol{S}}\bar{\boldsymbol{G}}$, where the term $\bar{\boldsymbol{S}}$ can be easily constructed based on the current local coordinates of the nodes by Eq. (7)/(16) while the term $\bar{\boldsymbol{G}}$ defined by Eq. (9)/(17) is a highly nonlinear function of the local coordinates which is hard to deduce for some choices of local CR frame (such as the polar decomposition approach). In our numerical examples, we use the high-accuracy complex step finite difference (CSFD) method [43-45] to obtain this term. The CSFD method for a scalar function is summarized below.

Consider the complex-differentiable function $f:\mathbb{C}\rightarrow\mathbb{C}$ perturbed about the nominal point $\bar{x}$ by $ih$ where $\bar{x}, h \in \mathbb{R}$ and $i = \sqrt{-1}$. A Taylor series expansion yields

$$f(\bar{x} + ih) = f(\bar{x}) + \left.\frac{\partial f}{\partial z}\right|_{z=\bar{x}} ih - \frac{h^2}{2!}\left.\frac{\partial^2 f}{\partial z^2}\right|_{z=\bar{x}} - \frac{ih^3}{3!}\left.\frac{\partial^3 f}{\partial z^3}\right|_{z=\bar{x}} + o\left(h^4\right) \tag{63}$$

If $f(\bar{x})$ is assumed to be real for all real $\bar{x}$, taking the imaginary part of the above expression yield the complex step finite difference scheme for the first-order approximation.

$$\frac{\partial f}{\partial z} = \frac{\operatorname{Im}\left\{f(\bar{x} + ih)\right\}}{h} + o\left(h^2\right) \approx \frac{\operatorname{Im}\left\{f(\bar{x} + ih)\right\}}{h} \tag{64}$$

The significate advantage of the CSFD method over traditional (real step-based) finite difference methods is that it obviates the subtractive cancellation problem for the first-order approximation, and an extremely small value of the perturbation step size can be applied (e.g. $h = 1\times10^{-20}$ or even smaller). The use of small difference step size enables the CSFD method to completely replace the analytic formulation within machine precision. As an simple example, the Matlab code snippet of using the CSFD method to obtain the first-order derivative of function $f(x) = \mathrm{e}^x + x^2 + 1$ at $x = 0.5$ is given below (with $h = 1\times10^{-50}$).

```
%% calFunction.m
function [f] = calFunction(x)
f=exp(x)+x^2+1;
end
```

```
%% main.m
h=1e-50;
x=0.5;
dfdx = imag(calFunction(x+1i*h))/h ;
%% output dfdx =2.64872127070013
```

## References

[1] G. Wempner, Finite elements, finite rotations and small strains of flexible shells, Int J Solids Struct, 5 (1969) 117-153.

[2] T. Belytschko, B.J. Hsieh, Non-linear transient finite element analysis with convected co-ordinates, Int J Numer Meth Eng, 7 (1973) 255-271.

[3] T. Belytschko, L. Schwer, M.J. Klein, Large displacement, transient analysis of space frames, Int J Numer Meth Eng, 11 (1977) 65-84.

[4] J.H. Argyris, H. Balmer, J.S. Doltsinis, P.C. Dunne, M. Haase, M. Kleiber, G.A. Malejannakis, H.P. Mlejnek, M. Müller, D.W. Scharpf, Finite element method — the natural approach, Comput Method Appl M, 17-18 (1979) 1-106.

[5] C.C. Rankin, F.A. Brogan, An Element Independent Corotational Procedure for the Treatment of Large Rotations, Journal of Pressure Vessel Technology, 108 (1986) 165-174.

[6] C.C. Rankin, B. Nour-Omid, The use of projectors to improve finite element performance, Computers & Structures, 30 (1988) 257-267.

[7] B. Nour-Omid, C.C. Rankin, Finite rotation analysis and consistent linearization using projectors, Comput Method Appl M, 93 (1991) 353-384.

[8] J.M. Battini, A non-linear corotational 4-node plane element, Mech Res Commun, 35 (2008) 408-413.

[9] H. Cho, H. Joo, S. Shin, H. Kim, Elastoplastic and contact analysis based on consistent dynamic formulation of co-rotational planar elements, Int J Solids Struct, 121 (2017) 103-116.

[10] H. Cho, J. Kwak, S. Shin, N. Lee, S. Lee, Flapping-Wing Fluid–Structural Interaction Analysis Using Corotational Triangular Planar Structural Element, AIAA J., 54 (2016) 2265-2276.

[11] M.A. Crisfield, G.F. Moita, A unified co-rotational framework for solids, shells and beams, Int J Solids Struct, 33 (1996) 2969-2992.

[12] M. Mostafa, M.V. Sivaselvan, On best-fit corotated frames for 3D continuum finite elements, Int J Numer Meth Eng, 98 (2014) 105-130.

[13] H. Cho, S. Shin, J.J. Yoh, Geometrically nonlinear quadratic solid/solid-shell element based on consistent corotational approach for structural analysis under prescribed motion, Int J Numer Meth Eng, 112 (2017) 434-458.

[14] H. Cho, H. Kim, S. Shin, Geometrically nonlinear dynamic formulation for three-dimensional co-rotational solid elements, Comput Method Appl M, 328 (2018) 301-320.

[15] M.A. Crisfield, A consistent co-rotational formulation for non-linear, three-dimensional, beam-elements, Comput Method Appl M, 81 (1990) 131-150.

[16] C. Pacoste, A. Eriksson, Beam elements in instability problems, Comput Method Appl M, 144 (1997) 163-197.

[17] J.-M. Battini, C. Pacoste, Co-rotational beam elements with warping effects in instability problems, Comput Method Appl M, 191 (2002) 1755-1789.

[18] T. Macquart, S. Scott, P. Greaves, P.M. Weaver, A. Pirrera, Corotational Finite Element Formulation for Static Nonlinear Analyses with Enriched Beam Elements, AIAA J., 58 (2020) 2276-2292.

[19] M. Aguirre, S. Avril, An implicit 3D corotational formulation for frictional contact dynamics of beams against rigid surfaces using discrete signed distance fields, Comput Method Appl M, 371 (2020) 113275.

[20] M.A. Crisfield, U. Galvanetto, G. Jeleni, Dynamics of 3-D co-rotational beams, Comput Mech, 20 (1997) 507-519.

[21] S. Chhang, J.-M. Battini, M. Hjiaj, Energy-momentum method for co-rotational plane beams: A comparative study of shear flexible formulations, Finite Elem Anal Des, 134 (2017) 41-54.

[22] C. Pacoste, Co-rotational flat facet triangular elements for shell instability analyses,

Comput Method Appl M, 156 (1998) 75-110.

[23] E. Gal, R. Levy, Geometrically nonlinear analysis of shell structures using a flat triangular shell finite element, Arch Comput Method E, 13 (2006) 331-388.

[24] J.M. Battini, A modified corotational framework for triangular shell elements, Comput Method Appl M, 196 (2007) 1905-1914.

[25] F. Caselli, P. Bisegna, Polar decomposition based corotational framework for triangular shell elements with distributed loads, Int J Numer Meth Eng, 95 (2013) 499-528.

[26] Z. Li, Y. Xiang, B.A. Izzuddin, L. Vu-Quoc, X. Zhuo, C. Zhang, A 6-node co-rotational triangular elasto-plastic shell element, Comput Mech, 55 (2015) 837-859.

[27] Y.Q. Tang, Y.P. Liu, S.L. Chan, Element-Independent Pure Deformational and Co-Rotational Methods for Triangular Shell Elements in Geometrically Nonlinear Analysis, International Journal of Structural Stability and Dynamics, 18 (2018) 1850065.

[28] F.S. Almeida, A.M. Awruch, Corotational nonlinear dynamic analysis of laminated composite shells, Finite Elem Anal Des, 47 (2011) 1131-1145.

[29] J. Shi, Z. Liu, J. Hong, A New Rotation-Free Shell Formulation Using Exact Corotational Frame for Dynamic Analysis and Applications, Journal of Computational and Nonlinear Dynamics, 13 (2018).

[30] V.V. Kuznetsov, S.V. Levyakov, Phenomenological invariant-based finite-element model for geometrically nonlinear analysis of thin shells, Comput Method Appl M, 196 (2007) 4952-4964.

[31] P. Khosravi, R. Ganesan, R. Sedaghati, Corotational non-linear analysis of thin plates and shells using a new shell element, Int J Numer Meth Eng, 69 (2007) 859-885.

[32] H. Sung, H. Kim, J. Choi, H. Kim, C. Li, M. Cho, Structural design of soft robotics using a joint structure of photoresponsive polymers, Smart Materials and Structures, 29 (2020) 055032.

[33] L. Zhang, K. Dong, M. Lu, H. Zhang, A wrinkling model for pneumatic membranes and the complementarity computational framework, Comput Mech, 65 (2019) 119-134.

[34] B.A. Izzuddin, An enhanced co-rotational approach for large displacement analysis of plates, Int J Numer Meth Eng, 64 (2005) 1350-1374.

[35] Z.X. Li, B.A. Izzuddin, L. Vu-Quoc, A 9-node co-rotational quadrilateral shell element, Comput Mech, 42 (2008) 873-884.

[36] Z.X. Li, T.Z. Li, L. Vu-Quoc, B.A. Izzuddin, X. Zhuo, Q. Fang, A nine-node corotational curved quadrilateral shell element for smooth, folded, and multishell structures, Int J Numer Meth Eng, 116 (2018) 570-600.

[37] G. Wang, Z. Qi, J. Xu, A high-precision co-rotational formulation of 3D beam elements for dynamic analysis of flexible multibody systems, Comput Method Appl M, 360 (2020) 112701.

[38] C.A. Felippa, B. Haugen, A unified formulation of small-strain corotational finite elements: I. Theory, Comput Method Appl M, 194 (2005) 2285-2335.

[39] P. Areias, J. Garção, E.B. Pires, J.I. Barbosa, Exact corotational shell for finite strains and fracture, Comput Mech, 48 (2011) 385-406.

[40] M. Ritto-Corrêa, D. Camotim, Work-conjugacy between rotation-dependent moments and finite rotations, Int J Solids Struct, 40 (2003) 2851-2873.

[41] L.H. Teh, “Work-conjugacy between rotation-dependent moments and finite rotations” by Manuel Ritto-Correa and Dinar Camotim [Vol. 40, No. 11, pp. 2851–2873], Int J Solids Struct, 40 (2003) 5861-5863.

[42] M. Ritto-Corrêa, D. Camotim, Reply to: Dr. Lip Teh's discussion on "Work-conjugacy between rotation-dependent moments and finite rotations" [Vol. 40, No. 11, pp. 2851–2873], Int J Solids Struct, 40 (2003) 6211-6214.

[43] W. Squire, G. Trapp, Using Complex Variables to Estimate Derivatives of Real Functions, Siam Rev, 40 (1998) 110-112.

[44] J.N. Lyness, C.B. Moler, Numerical Differentiation of Analytic Functions, Siam J Numer Anal, 4 (1967) 202-210.

[45] J.N. Lyness, Differentiation formulas for analytic functions, Math Comput, 22 (1968) 352-352.

[46] I. Romero, A comparison of finite elements for nonlinear beams: the absolute nodal coordinate and geometrically exact formulations, Multibody Syst Dyn, 20 (2008) 51-68.

[47] J.C. Simo, L. Vu-Quoc, A three-dimensional finite-strain rod model. part II: Computational aspects, Comput Method Appl M, 58 (1986) 79-116.

[48] R.H. Macneal, R.L. Harder, A proposed standard set of problems to test finite element accuracy, Finite Elem Anal Des, 1 (1985) 3-20.

[49] B.A. Izzuddin, Y. Liang, Bisector and zero-macrospin co-rotational systems for shell elements, Int J Numer Meth Eng, 105 (2016) 286-320.

[50] K.Y. Sze, X.H. Liu, S.H. Lo, Popular benchmark problems for geometric nonlinear analysis of shells, Finite Elem Anal Des, 40 (2004) 1551-1569.

[51] J. Argyris, An excursion into large rotations, Comput Method Appl M, 32 (1982) 85-155.

[52] R.A. Spurrier, Comment on " Singularity-Free Extraction of a Quaternion from a Direction-Cosine Matrix", J Spacecraft Rockets, 15 (1978) 255-255.

**Declaration of interests**

☒ The authors declare that they have no known competing financial interests or personal relationships that could have appeared to influence the work reported in this paper.

☐The authors declare the following financial interests/personal relationships which may be considered as potential competing interests: